\documentclass[superscriptaddress,aps,prb,twocolumn]{revtex4}
\usepackage{amsmath}
\usepackage{epsfig}
\usepackage{graphicx}
\usepackage{multirow}
\usepackage{array}
\usepackage{inputenc}
\usepackage{color}
\usepackage{tabularx}
\usepackage{textcomp}

\newcommand{\przr}{Pr$_2$Zr$_2$O$_7$}
\newcommand{\prsn}{Pr$_2$Sn$_2$O$_7$}
\newcommand{\prir}{Pr$_2$Ir$_2$O$_7$}
\newcommand{\prhf}{Pr$_2$Hf$_2$O$_7$}

\newcommand{\ndzr}{Nd$_2$Zr$_2$O$_7$}
\newcommand{\hoti}{Ho$_2$Ti$_2$O$_7$}
\newcommand{\tbti}{Tb$_2$Ti$_2$O$_7$}

\newcommand{\dyti}{Dy$_2$Ti$_2$O$_7$}

\newcommand{\pr}{Pr$^{3+}$}

\begin{document}
\author{S. Petit}
\email[]{sylvain.petit@cea.fr}
\affiliation{Laboratoire L\'eon Brillouin, CEA, CNRS, Universit\'e Paris-Saclay, CEA-Saclay, F-91191 Gif-sur-Yvette, France}
\author{E. Lhotel}
\email[]{elsa.lhotel@neel.cnrs.fr}
\affiliation{Institut N\'eel, CNRS and Univ. Grenoble Alpes, F-38042 Grenoble, France}
\author{S. Guitteny}
\affiliation{Laboratoire L\'eon Brillouin, CEA, CNRS, Universit\'e Paris-Saclay, CEA-Saclay, F-91191 Gif-sur-Yvette, France}
\author{O. Florea}
\affiliation{Institut N\'eel, CNRS and Univ. Grenoble Alpes, F-38042 Grenoble, France}
\author{J. Robert} 
\affiliation{Institut N\'eel, CNRS and Univ. Grenoble Alpes, F-38042 Grenoble, France}
\author{P. Bonville}
\affiliation{SPEC, CEA, CNRS, Universit\'e Paris-Saclay, CEA-Saclay, F-91191 Gif-sur-Yvette, France}
\author{I. Mirebeau}
\affiliation{Laboratoire L\'eon Brillouin, CEA, CNRS, Universit\'e Paris-Saclay, CEA-Saclay, F-91191 Gif-sur-Yvette, France}
\author{J. Ollivier}
\affiliation{Institut Laue Langevin, F-38042 Grenoble, France}
\author{H. Mutka}
\affiliation{Institut Laue Langevin, F-38042 Grenoble, France}
\author{E. Ressouche}
\affiliation{INAC, CEA and Univ. Grenoble Alpes, CEA Grenoble, F-38054 Grenoble, France}
\author{C. Decorse}
\affiliation{ICMMO, Universit\'e Paris-Sud, F-91405 Orsay, France}
\author{M. Ciomaga Hatnean}
\affiliation{Department of Physics, University of Warwick, Coventry, CV4 7AL, United Kingdom}
\author{G. Balakrishnan}
\affiliation{Department of Physics, University of Warwick, Coventry, CV4 7AL, United Kingdom}

\title{Antiferro-quadrupolar correlations in the quantum spin ice candidate \przr}

\begin{abstract}
We present an experimental study of the quantum spin ice candidate pyrochlore coumpound \przr\ by means of magnetization measurements, specific heat and neutron scattering up to 12 T and down to 60 mK. When the field is applied along the $[111]$ and $[1\bar{1}0]$ directions, ${\bf k}=0$ field induced structures settle in. We find that the ordered moment rises slowly, even at very low temperature, in agreement with macroscopic magnetization. Interestingly, for $H \parallel [1\bar{1}0]$, the ordered moment appears on the so called $\alpha$ chains only. The spin excitation spectrum is essentially {\it inelastic} and consists in a broad flat mode centered at about 0.4 meV with a magnetic structure factor which resembles the spin ice pattern. For $H \parallel [1\bar{1}0]$ (at least up to 2.5 T), we find that a well defined mode forms from this broad response, whose energy increases with $H$, in the same way as the temperature of the specific heat anomaly. We finally discuss these results in the light of mean field calculations and propose a new interpretation where quadrupolar interactions play a major role, overcoming the magnetic exchange. In this picture, the spin ice pattern appears shifted up to finite energy because of those new interactions. We then propose a range of acceptable parameters for \przr\, that allow to reproduce several experimental features observed under field. With these parameters, the actual ground state of this material would be an antiferroquadrupolar liquid with spin-ice like excitations. 
\end{abstract}

\maketitle
\section{Introduction}

The concept of geometrical frustration has attracted much attention in physics. It covers a wide variety of situations where a local configuration, stabilized by a given scheme of interactions, cannot extend simply over the whole system. Numerous examples can be found in pentagonal or icosahedral lattices, metallic binary alloys, liquid crystals, the bistable states of metal organic networks, the packing of molecules on triangular lattices, among others \cite{Mosseri99}. 

\begin{figure}[t]
\includegraphics[width=5cm]{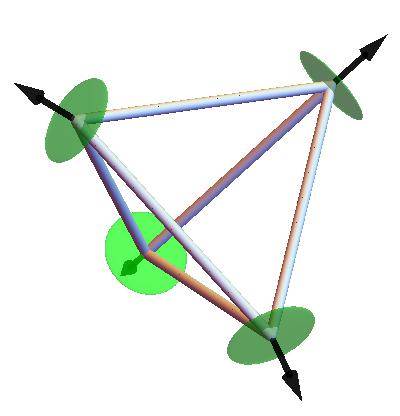}
\caption{\label{Fig1} Local $\vec{z}_i$ anisotropy axes in a tetrahedron of the pyrochlore lattice. The green disks represent the local $xy$ planes. For ions located at $(1/4, 1/4, 1/2)$ and related symmetry positions, $\vec{z}_i=(1,1,-1)/\sqrt{3}$, for $(1/2, 1/2, 1/2)$, $\vec{z}_i=(-1,-1,-1)/\sqrt{3}$, for $(1/2, 1/4, 1/4)$, $\vec{z}_i=(-1,1,1)/\sqrt{3}$ and $(1/4, 1/2, 1/4)$, $\vec{z}_i=(1,-1,1)/\sqrt{3}$.}
\end{figure}

In condensed matter physics, the archetype of geometrical frustration in three dimensions is the problem of Ising spins that reside on the vertices of the pyrochlore lattice, built from corner sharing tetrahedra \cite{Lacroix,Gardner10,Gingras14}. If the spins are constrained to lie along the local axes which link the center of a tetrahedron to its summits (denoted hereafter $\vec{z}_i$, see Figure \ref{Fig1}), and experience ferromagnetic interactions (for example due to the magnetic dipolar interaction), a disordered highly degenerate ground state, the spin ice state, develops at low temperature \cite{Harris97,Ramirez99,denHertog00,Castelnovo08}. The nearest-neighbor ferromagnetic coupling favors local configurations where in each tetrahedron, two spins point into and two out of the center (``2-in$-$2-out'' configurations), forming a magnetic analog of the water ice. One of the clear proof of this physics came with the observation of magnetic diffuse scattering in \hoti\, and \dyti, characterized by arm-like features in reciprocal space along with specific bow tie singularities also called pinch points \cite{Fennell09,Morris09}, in excellent agreement with theoretical calculations \cite{Isakov04,Henley05,Henley10}.

While thermal heating naturally melts the spin ice, the possibility that quantum fluctuations might also melt spin ice is a topical and fascinating issue. Provided that transverse terms, as opposed to the ``classical'' ferromagnetic interaction between Ising spins, are not too large, several theoretical works have claimed that the physics can be described by an emergent electrodynamics with new deconfined particles \cite{Hermele04,Benton12,Gingras14}.  
Recently, several theoretical studies \cite{Onoda10,Onoda11,Lee12} have proposed the \pr\, based pyrochlore magnets like for instance \przr\, as good candidates. A light rare-earth is indeed expected to enhance transverse interactions because of a large overlap between 4f and oxygen orbitals.

Experiments on \prsn \cite{Matsuhira02, Princep13}, \przr \cite{Matsuhira09,Kimura13,Monica14}, \prir \cite{Nakatsuji06} and more recently \prhf \cite{Sibille16} have shown that the \pr\, moment has a strong Ising character, described by a non-Kramers magnetic doublet. As in spin ice, no magnetic long range ordering is observed down to dilution temperature, and magnetic specific heat shows a broad peak at about 2 K \cite{Lutique04,Matsuhira09, Kimura13, Sibille16, Nakatsuji06, Zhou08}, similar to what is observed in the classical spin ice \dyti. 

At $T \approx 0.1$~K, neutron scattering measurements in \przr\, reveal that fluctuating magnetic correlations develop, with a very weak elastic component representing less than 10\% of the response \cite{Kimura13}. Their wave vector dependence shows features similar to the spin ice pattern, yet the pinch points appear broadened. These results were interpreted as the evidence of quantum dynamics in a new class of spin ice system.

Nevertheless, in \przr\, and \prhf\, the Curie-Weiss temperature inferred from magnetic susceptibility is negative \cite{Matsuhira09, Kimura13, Monica14, Sibille16}, thus indicating antiferromagnetic interactions, which is a priori not consistent with the spin ice picture. In addition, the fact that most of the neutron scattering signal in \przr\, has an inelastic character calls for peculiar spin dynamics, different from conventional spin-ice. These issues are still to be answered and a key ingredient to clarify them may be the quadrupolar degrees of freedom.
Indeed, the latter are known to play an important role in the physics of non-Kramers ions such as \pr. Quadrupole (and even multipole) interactions in rare-earth magnets are naturally induced by superexchange and electrostatics \cite{Santini09,Wolf68,Rau16} and were put forward as an essential ingredient to describe \przr\, from a  theoretical point of view\cite{Onoda10}. 

The aim of the present work is to shed light on the peculiar ground state of \przr. First, we address the non-Kramers ion (like \pr) specificities in the context of pyrochlore magnets. We especially point out the need for special care to interprete neutron data because the moment of non-Kramers doublets has different properties from usual magnetic moments. With this result in hand, we explore the ground state and magnetic excitations in \przr\, by means of magnetization, specific heat, neutron diffraction and inelastic neutron scattering.  In particular, we investigate the field induced properties, in macroscopic and neutron scattering measurements. We determine the magnetic field induced structure, and show the existence of a magnetic excitation whose energy is shifted by the magnetic field. 

Using a mean field treatment of the minimal Hamiltonian widely accepted in the literature for these materials \cite{Gingras14}, it emerges that these observations can be understood by considering that the dominant coupling at play is an effective quadrupolar interaction and {\it not} the ``classical'' ferromagnetic dipolar one as expected in spin ice. We show that effective quadrupolar interactions stabilize at this level of approximation, and for moderate positive or negative values of the interactions between Ising spins, an ``all-in$-$all-out'' quadrupolar phase reminiscent of the antiferro-quadrupolar Higgs phase found in more elaborate theories \cite{Lee12}. From this analysis and the comparison with the set of experiments, we propose a range of acceptable parameters for \przr. We conclude that the actual ground state of this material supports antiferroquadrupolar correlations. 

\section{Pyrochlore magnets and Non-Kramers ions}
\label{sectionnk}

\subsection{Crystal electric field}

In pyrochlore magnets, the crystal electric field Hamiltonian ${\cal H}_{\mbox{CEF}}$ is of fundamental importance as it determines the properties and symmetries of the lowest on site energy states. In \pr\ based systems, some studies have modeled this crystal field Hamiltonian by taking into account the set of electronic multiplets \cite{Princep13, Sibille16, Bonville}. Yet, for the sake of simplicity, we consider here the ground multiplet $J=4$ only and write: ${\cal H}_{\mbox{CEF}} = \sum_{m,n} B_{nm} O_{nm}$ where the $O_{nm}$ are the Wybourne operators \cite{wybourne}. The quantization axes are the $\vec{z}_i$ axes (black arrows in Figure \ref{Fig1}). 
The $B_{nm}$ coefficients have been determined in Ref. \citenum{Kimura13} and revisited in Ref. \citenum{Bonville} (see also Appendix \ref{appendix_CF}).  In this approach, the CEF ground state is a non-Kramers doublet $|\uparrow,\downarrow \rangle$, well separated from the excited levels, with the general form (in the $|J_z = -J,..,J \rangle$ space):
\begin{eqnarray*}
|\uparrow     \rangle &=& \left(a,0,0,b,0,0,c,0,0 \right) \\
|\downarrow \rangle &=& \left(0,0,c,0,0,-b,0,0,a \right)
\end{eqnarray*}
The first excited state is a singlet:
\begin{eqnarray*}
|1     \rangle &=& \left(0, -d, 0, 0, e, 0, 0, d, 0 \right)
\end{eqnarray*}
The normalization condition assumes $a^2+b^2+c^2=1$ and $2d^2+e^2=1$. Using this explicit formulation, it is possible to calculate the projection of the magnetic moment $\vec{J}$ onto the 2$\times$2 subspace spanned by $|\uparrow,\downarrow \rangle$:
\begin{eqnarray}
J_x & = &  0, \nonumber \\
J_y & = &  0, \nonumber \\
J_z & = &  \left(
\begin{array}{cc} 
-\mu & 0 \\
0     & \mu
\end{array}
\right)
\label{dipoles}
\end{eqnarray}
with $\mu= 4 a^2 + b^2 - 2 c^2$.
In other words, the components of $\vec{J}$ can be written using an effective anisotropic $g$ factor defined within the ground-state doublet: 
\begin{eqnarray*}
 g_{\perp} &=& g_x = g_y \equiv 0 \\
 g_{\parallel} & = & 2 g_{\rm J} \mu
\end{eqnarray*}
It is also possible to calculate the quadrupolar operators. Their projection onto the subspace spanned by $|\uparrow \downarrow \rangle$ leads to:
\begin{eqnarray}
J_+^2+J_-^2 & = &  2\left(-20~b^2 + 8 \sqrt{7}~a~c\right)~\left(
\begin{array}{cc} 
0 & 1/2 \\
1/2     & 0
\end{array}
\right) \nonumber \\
J_x J_y + J_y J_x & = & 2\left( -10b^2 - 4  \sqrt{7}~a~c\right)~ \left(
\begin{array}{cc} 
0 & i/2 \\
-i/2 & 0
\end{array}
\right) \nonumber \\
J_x J_z + J_z J_x & = & -18 \sqrt{2}~b~c~ \left(
\begin{array}{cc} 
0 & 1/2 \\
1/2     & 0
\end{array}
\right) \nonumber \\
J_y J_z + J_z J_y & = & -18 \sqrt{2}~b~c~ \left(
\begin{array}{cc} 
0 & i/2 \\
-i/2 & 0
\end{array}
\right) 
\label{quadrupoles}
\end {eqnarray}
Note that the fifth quadrupolar operator $3J_z^2 - J(J+1)$ is proportionnal to the identity in this subspace and thus not relevant. As shown by the above matrix representation of Eq. (\ref{dipoles}), it is clear that fluctuations within the ground doublet cannot be induced by magnetic exchange since $\langle \uparrow | \vec{J} | \downarrow \rangle \equiv 0$. This is the key property of non-Kramers doublets. However, Eq. (\ref{dipoles}) and (\ref{quadrupoles}) form together the set of Pauli matrices of a pseudo spin 1/2, $\vec{\sigma}=(\sigma^x,\sigma^y,\sigma^z)$. Those pseudo spins reside on the pyrochlore lattice sites. The $z$ components describe the Ising magnetic moments pointing along the $\vec{z}_i$ axes and the $x$ and $y$ components (hence $\sigma^+_i$ and $\sigma^-_i$) correspond to the quadrupolar ``degrees of freedom". Fluctuations within the ground doublet are thus naturally reintroduced by those degrees of freedom.

\subsection{General Hamiltonian}

On this ground, a general Hamiltonian has been proposed in Ref. \citenum{Rossprx11,Savary12} and adapted to the case of non-Kramers ions in Ref. \citenum{Curnoe07,Curnoe14,Onoda10,Onoda11,Lee12}. It is bilinear in terms of the local components of pseudo spins 1/2: 
\begin{eqnarray}
{\cal H} = \frac{1}{2} \sum_{<i,j>} {\cal J}^{zz} {\sf \sigma}^z_i {\sf \sigma}^z_j + \sum_i (g_{\parallel} \mu_{\rm B} \vec{z}_i   \cdot \vec{h})~{\sf \sigma}^z_i  \nonumber \\
+ \frac{1}{2} \sum_{<i,j>} -{\cal J}^{\pm} \left(\sigma^+_i \sigma^-_j + \sigma^-_i \sigma^+_j\right) \nonumber \\
+\frac{1}{2} \sum_{<i,j>} {\cal J}^{\pm\pm} \left(\gamma_{ij} \sigma^+_i \sigma^+_j + \gamma^*_{ij}\sigma^-_i \sigma^-_j \right) 
\label{h}
\end{eqnarray}
The $\gamma_{ij}$ parameter is defined in Ref. \citenum{Rossprx11}. ${\cal J}^{\pm}$ and ${\cal J}^{\pm\pm}$ are effective quadrupolar exchange terms, compatible with the local symmetry of the rare earth. Note that information on the actual microscopic interactions between the 4f \pr\, electrons is lost through the projection into the ground doublets \cite{Rau16}. From a physical point of view, ${\cal J}^{\pm}$ and ${\cal J}^{\pm\pm}$ promote quadrupolar states with orientations of $\sigma$ perpendicular to the local $\vec{z}$ axis. They correspond to so-called transverse or quantum terms, in contrast to the Ising coupling ${\cal J}^{zz}$. The latter couples the local $z$ components only and derives from the combination of the original exchange coupling ${\cal J}$ and of the dipolar interaction truncated to nearest neighbors:
\begin{equation*}
{\cal J}^{zz}= \frac{g_{\parallel}^2}{g_{\rm J}^2}~\left(\frac{-{\cal J} + 5 {\cal D}}{3} \right)
\end{equation*}
with ${\cal D}= \dfrac{\mu_o (g_{\rm J} \mu_{\rm B})^2}{4\pi r_{\rm nn}^3}$ ($r_{\rm nn}$ is the nearest neighbor distance between rare-earth ions).  When it is positive, i.e. when the dipolar term overcomes the antiferromagnetic exchange, the spin-ice state develops, while in the opposite situation, the ``all-in$-$all-out'' antiferromagnetic state is expected \cite{Bramwell01}. 

Note that a magnetic field $\vec{h}$ would couple to $\sigma^z$ only, while a strain (or distortion) would couple to the quadrupolar electronic degrees of freedom; this would be taken into account by an effective ``strain'' field $v_i$ coupled to $\sigma^+$ and $\sigma^-$:
\begin{eqnarray}
{\cal H}_v  = {\cal H} + \sum_i v_i ~{\sf \sigma}^+_i + v_i^* ~{\sf \sigma}^-_i
\label{h2}
\end{eqnarray}

\subsection{Consequences for the interpretation of magnetic measurements}

Magnetic measurements, especially macroscopic magnetization or neutron scattering, are however not sensitive to the pseudo spin $\sigma$ but to the actual magnetic moment operators $\vec{J}$. This has consequences when interpreting the data. To illustrate this point, we determine the formal expression of the dynamical spin-spin correlation function $S(Q,\omega)$ measured by neutron scattering.

In a classical picture, the ground state $|\Phi_G \rangle$ of the above Hamiltonian (\ref{h2}) can be described as a state where on each site of the pyrochlore lattice, the expectation value of the pseudo spin $\vec{\sigma}=(\sigma_x,\sigma_y, \sigma_z)$ is oriented in the direction specified by {\it local} spherical angles $\theta_i$ and $\phi_i$: $\theta_i$ defines the polar angle relative to the local CEF axes; $\phi_i$ is the angle within the $xy$ plane (green disks in Figure \ref{Fig1}):
\begin{equation*}
\begin{array}{ccc}
|\Phi_G \rangle & = & | \phi_{G,1}~...~\phi_{G,i}~ ...~ \phi_{G,N} \rangle
\end{array}
\end{equation*}
where $N$ is the (infinite) number of sites. Those angles depend on the parameters of the Hamiltonian but it is not necessary to specify them at this step. Then, as expected for instance in the Random Phase Approximation (RPA) or spin wave approximation, the lowest energy excited states $|\Phi_1 \rangle$, with energy $E_1$ above the ground state, should contain one flip of the pseudo spin, possibly delocalized over the lattice. $|\Phi_1 \rangle$ is thus constructed as:
\begin{equation*}
|\Phi_1 \rangle = \sum_i  ~ C_i~| \phi_{G,1}~...~\phi_{1,i}~ ...~ \phi_{G,N} \rangle
\end{equation*}
where $|\phi_{1,i}\rangle$ describes a flip of the pseudo spin $\sigma$ at site $i$. The values of the $C_i$ coefficients depend on the Hamiltonian and remain to be determined. 

At low temperature, keeping the ground and first excited states, $S(Q,\omega)$ can be approximated by (see Appendix \ref{appendix_Sq} for details):
\begin{eqnarray*}
S(Q,\omega=0)        &\approx & \mu^2  |\sum_{i} e^{iQ R_i} \cos \theta_i ~\vec{z}_{\perp,i} |^2 \\
S(Q,\omega=E_1) &\approx & \mu^2   |\sum_{i} C_i~e^{iQ R_i} ~ e^{i\phi_i} \sin \theta_i ~\vec{z}_{\perp,i} |^2
\end{eqnarray*}
hence to an elastic contribution at $\omega=0$, and an inelastic one at $\omega=E_1$. The symbol $\perp$ indicates that one must consider the components perpendicular to the scattering wavevector $Q$. 

\subsubsection{Magnetic states}

It is first instructive to examine the case of ``magnetic" states ($\theta_i =0, \pi)$, where the pseudo-spins point along the $\vec{z}$ directions. The {\it elastic} contribution $S(Q,\omega=0)$ writes 
\begin{equation*}
S(Q,\omega=0) \approx  \mu^2  |\sum_{i} e^{iQ R_i} \epsilon_i ~\vec{z}_{\perp,i} |^2
\end{equation*}
with $\epsilon_i =\pm 1$ (depending on the values of $\theta_i$). Spin ice corresponds to the case where, in each tetrahedron, there are two sites with $\theta_i=0$ and two with $\theta_i=\pi$. Then, $S(Q,\omega=0)$ has arm like features along $(00\ell)$ and $(111)$ with pinch points at $(002)$, and $(111)$ positions in reciprocal space \cite{Henley05}. 
In contrast, it is clear from the above formula that the non-Kramers nature of the moments cancels the {\it inelastic} contribution: $S(Q,\omega=\Delta)=0$. 

\subsubsection{Quadrupolar states}

In the case of quadrupolar states $\theta_i= \pi/2$, the opposite situation is obtained. The {\it elastic} contribution is zero, as expected since the ground state is not magnetic, while the {\it inelastic} contribution $S(Q,\omega=\Delta)$ is finite. The dynamical part becomes observable because it corresponds to {\it magnetic} transitions from the ground state. Further, provided $C_i e^{i\phi_i} \sin \theta_i = \pm 1$ as the $\epsilon_i$ do in the case of spin ice, the spin ice pattern will appear shifted towards finite energy. We shall come back to this point in the discussion presented in section \ref{discuss}. 

\vspace{1cm}
With these results in hand, which specify the context of our study, we now turn to the description of the experimental results.

\section{Experimental results}
\subsection{Crystal growth}

\begin{figure}[h]
\includegraphics[width=4cm]{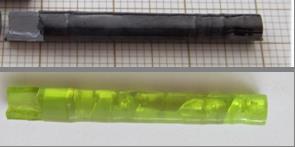}
\caption{\label{photo-pr} Picture of the single crystal, as grown (top) and annealed (bottom) used in the present study.}
\end{figure}

A single crystal was synthesized at the Physics Department of Warwick University from feed rods of \przr\, composition using the floating zone technique. The crystal growth was conducted in air, using a four-mirror xenon arc lamp optical furnace (CSI FZ-T-12000-X-VI VP, Crystal System Incorporated, Japan) \cite{Monica14, Monica15}. The as-grown crystal, dark-brown in colour, was annealed for two days in Ar (10\% H$_2$) flow at 1200 $^{\circ}$C and became bright green. This color change is associated, as suggested by Nakatsuji {\it et al}\, \cite{Nakatsuji06} with the modification of the oxidation state of Pr$^{4+}$ ions present in very small quantities in the dark-brown sample, to Pr$^{3+}$ ions (see Figure~\ref{photo-pr}).

The structural X-Ray analysis \cite{Monica14} points to a stoichiometry close to the ideal pyrochlore composition (2:2:7) and is similar to those published in Ref. \onlinecite{Koohpayeh14}. Small deformations of the Bragg peaks have nevertheless been observed by means of diffuse neutron scattering experiments, which correspond to a local volume variation at the Pr site of about 1 \textperthousand. These inhomogeneities, even small, could affect the magnetic properties, due to the sensitivity of non-Kramers doublets to local perturbations \cite{Gingras14, Blanchard12, Foronda15, Duijn05}. Further studies are ongoing to investigate in details these inhomogeneities and their consequences. 


\subsection{Macroscopic measurements}

\subsubsection{Experimental details}
Magnetization and specific heat measurements were performed on a single crystal of 14.24 mg. Its non regular shape prevented us from making accurate demagnetization measurements. The results are thus presented without demagnetization corrections. Nevertheless, it is expected that the demagnetization factor is in the same range for the three measured directions. 

Magnetization and ac susceptibility measurements were performed in the 85 mK - 4.2 K temperature range on a SQUID magnetometer equipped with a dilution refrigerator developed at the Institut N\'eel \cite{Paulsen01}. The magnetization was measured along the [111], [110] and [100] directions of the sample. 
Specific heat measurements were performed on a Quantum Design PPMS with a $^3$He option. In these experiments, the field was applied along the [110] direction. 

\subsubsection{Magnetic measurements}
\label{magnetization}
\begin{figure}[!]
\includegraphics[width=8cm]{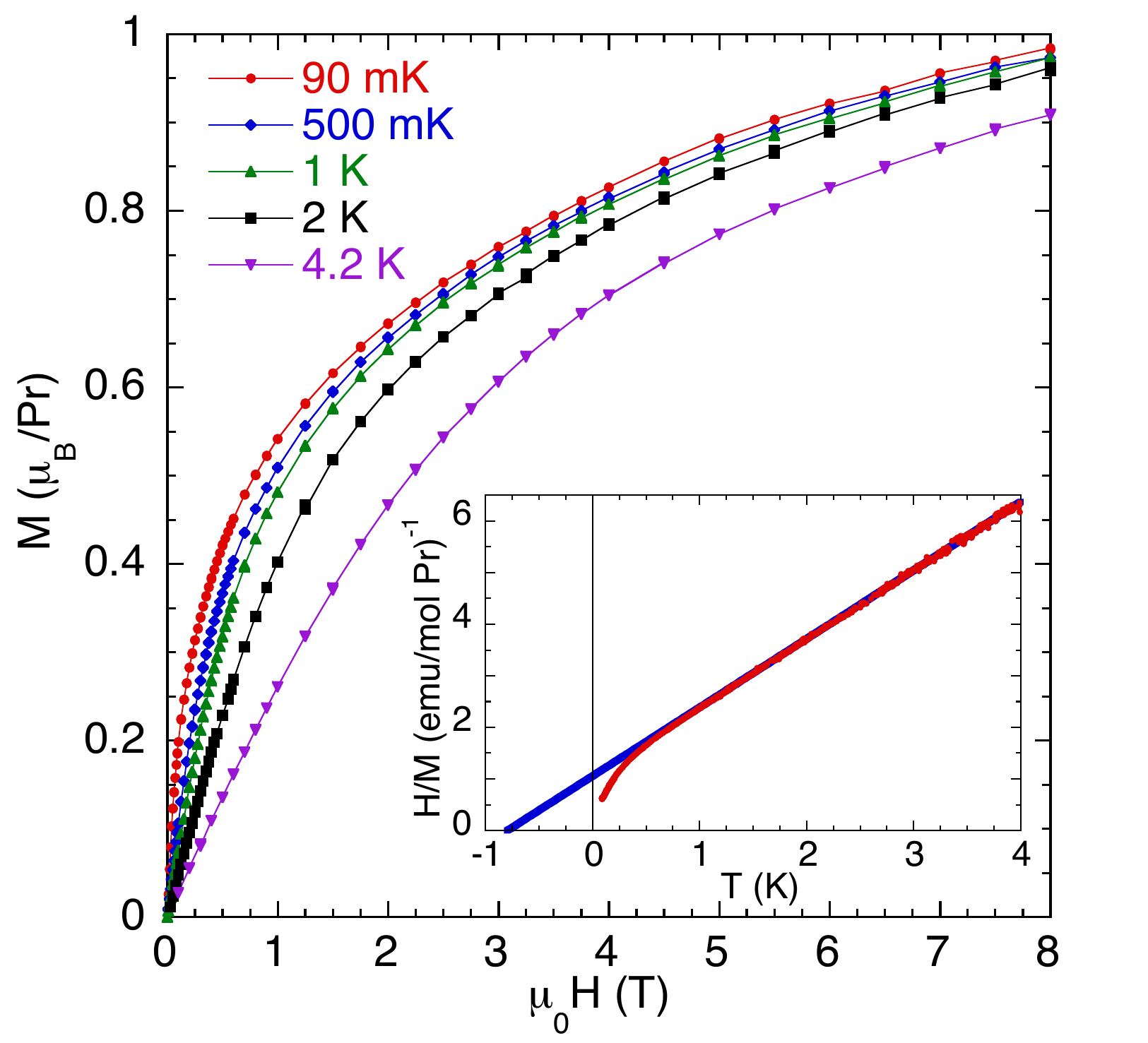}
\caption{\label{Fig_MH_X} $H \parallel [111]$: $M$ vs $H$ for several temperatures. Inset: $H/M$ vs $T$ in $\mu_0H=9$ mT. The line is a fit to the Curie-Weiss law between 1 and 4.2 K: $H/M=1.055+1.328 T$.}
\end{figure}

Magnetization as a function of temperature shows a continuous increase when the temperature decreases, and no signature of magnetic transition, nor zero field cooled - field cooled effects down to 90 mK. Note however that below 200 mK, equilibrium times become very long (about 500 s) which can lead to apparent hysteretic behavior. The susceptibility can be fitted to a Curie-Weiss law down to about 700 mK (see inset of Figure \ref{Fig_MH_X}) which gives an effective moment $\mu_{\rm eff}=2.45 \pm 0.02~\mu_{\rm B}$ and a Curie-Weiss temperature $\theta_{\rm CW}=-790 \pm 5$ mK. The value of the effective moment is in agreement with the value obtained in the CEF calculations in other Pr based pyrochlores taking into account the whole set of multiplets \cite{Princep13, Sibille16} as well as other magnetization measurements. The negative Curie-Weiss temperature is in the range of reported values for \przr, although some distribution is observed in the literature \cite{Matsuhira09, Kimura13, Monica14}, probably due to slightly different compositions between the samples \cite{Monica14}. 

\begin{figure}[!]
\includegraphics[width=8cm]{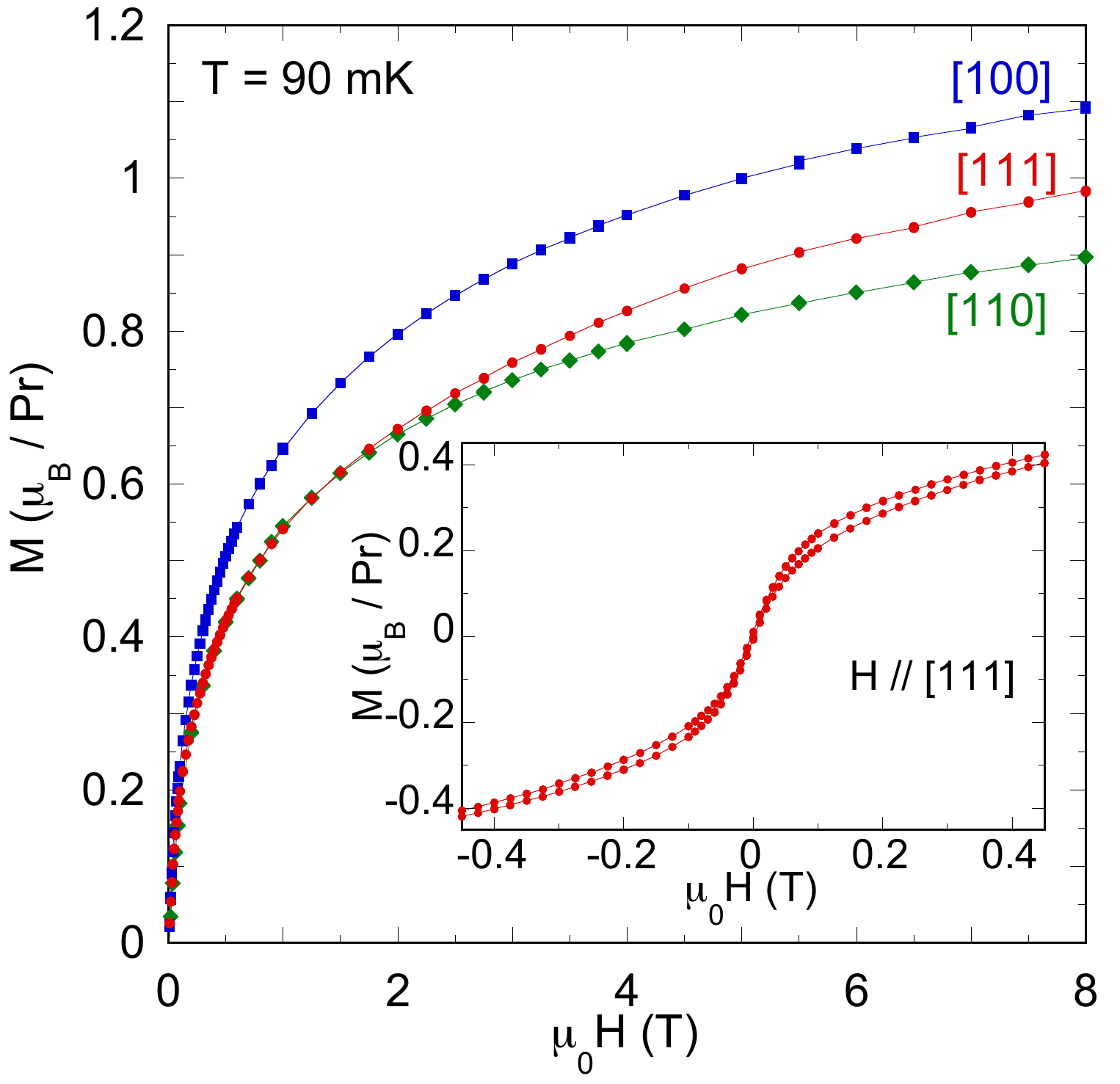}
\caption{\label{Fig_MH} $M$ vs $H$ at 90 mK for the field applied along the [111] (red dots), [110] (green triangles) and [100] (blue squares) directions of the sample. Inset: zoom on the low field part of the [111] magnetization showing a hysteretic behavior.}
\end{figure}

The magnetization curves at 90 mK for the field aligned along the three main directions of the cube are shown on Figure \ref{Fig_MH}. The magnetization is not fully saturated, even at 8 T. The reached magnetization is different along the three directions, as predicted for such Ising spins with a multiaxis anisotropy \cite{Harris98}. Nevertheless, the ratio between the obtained values are smaller than the expected ratio ($M_{\rm[100]}/M_{\rm[111]}=2/{\sqrt 3}$, $M_{\rm[110]}/M_{\rm[111]}=2/{\sqrt 6}$), suggesting that the apparent anisotropy is reduced compared to the case of classical Ising spins. In addition, the absolute values themselves are smaller than expected with an effective moment of 2.45 $\mu_{\rm B}$: for example $M_{\rm[111]}\approx 1~\mu_{\rm B}$ should be about 1.2 $\mu_{\rm B}$. The reason for this discrepancy between the saturated and effective moments is not understood at the moment. 

It is worth noting that a hysteretic behavior is observed at finite fields (see inset of Figure \ref{Fig_MH}), which reminds some bottleneck effects \cite{Vanvleck41}, but, in zero field, there is no remanent magnetization. 

\begin{figure}[!]
\includegraphics[width=8cm]{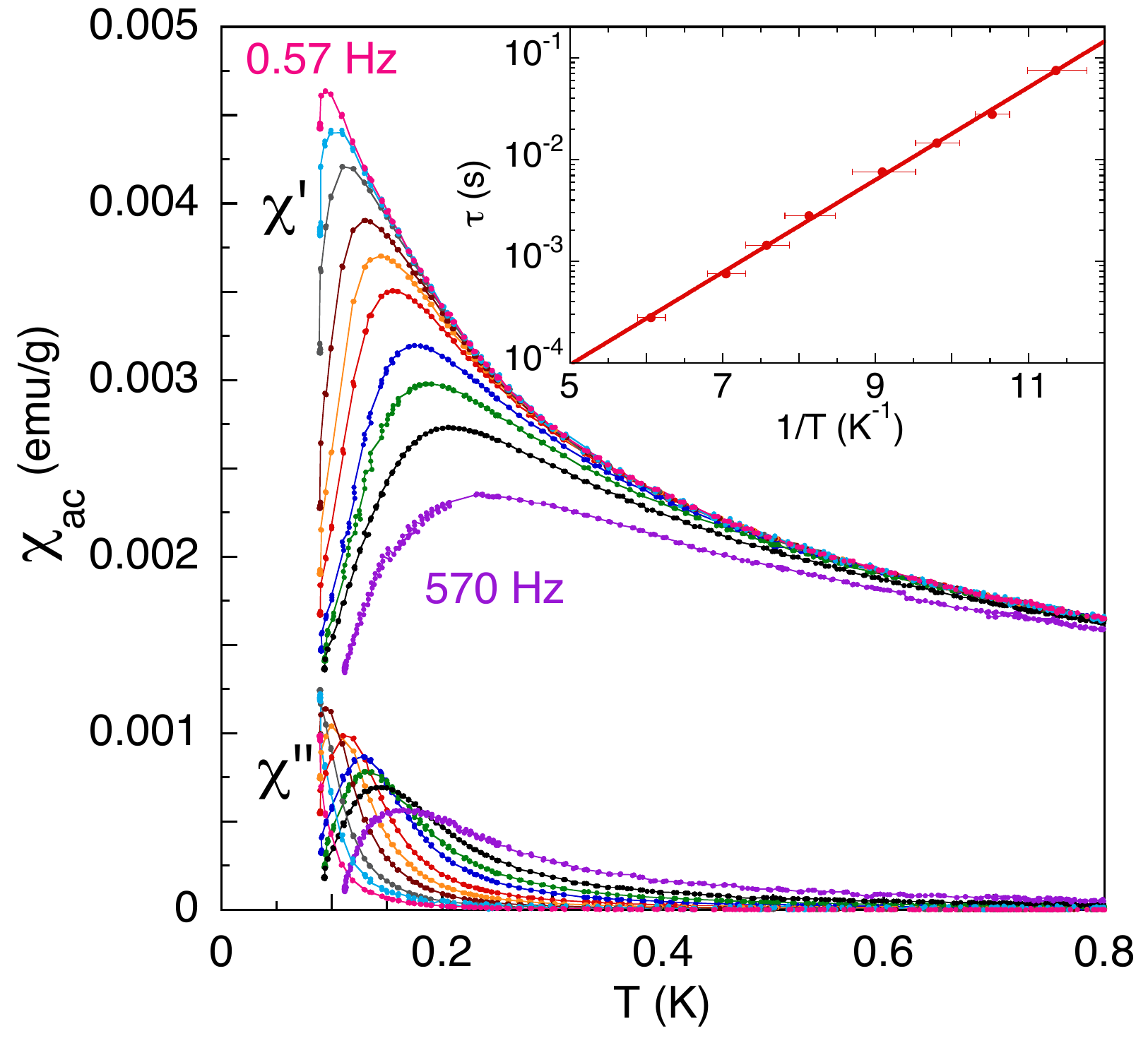}
\caption{\label{Fig_Xac} In-phase $\chi'$ and out-of-phase $\chi''$ parts of the ac susceptibility as a function of temperature, with $H_{\rm ac}=0.55$ mT, parallel to the [111] axis, for frequencies $f$ between 0.57 and 570 Hz. Inset: $\tau=1/2\pi f$ as a function of the inverse temperature of the $\chi''$ peak in a semilogarithmic scale. The line is a fit to the Arrhenius law: $\tau=5.1 \times 10^{-7} \exp(1.05/T)$. }
\end{figure}

Ac susceptibility measurements show a freezing as previously reported \cite{Matsuhira09, Kimura13}, which is characterized by a large signal in the dissipative part $\chi''$, and peaks in both $\chi'$ and $\chi''$ which move with frequency. The frequency dependence of the dissipative part of the susceptibility can be fitted by an Arrhenius law, as reported by Kimura et al. \cite{Kimura13}. Although in the same range, the obtained energy barrier, about 1 K (see inset of Figure \ref{Fig_Xac}), is smaller while the characteristic time $\tau_0\approx 5\times 10^{-7}$ s is larger. 

\subsubsection{Specific heat}
\begin{figure}[t]
\includegraphics[width=8cm]{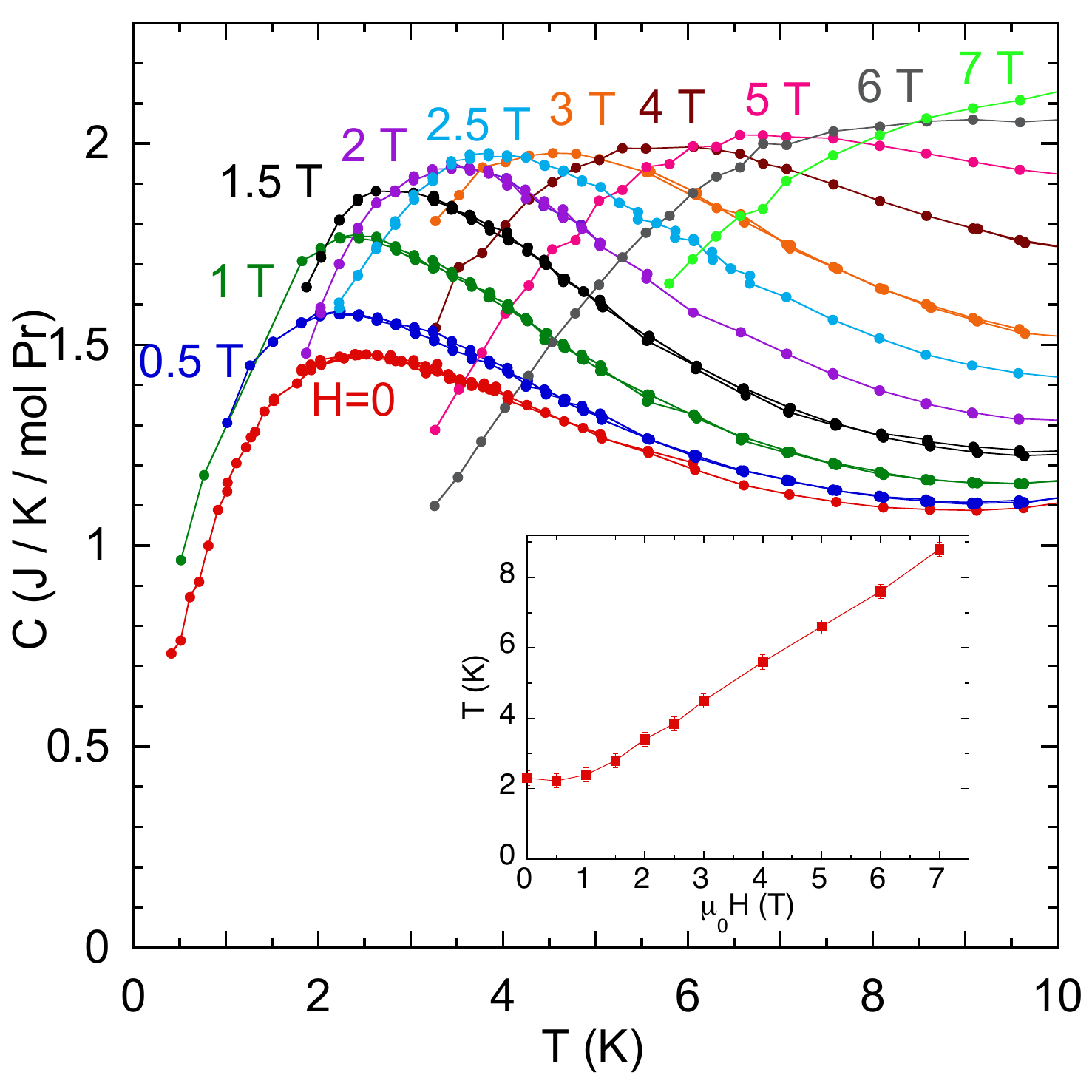}
\caption{\label{Fig_CT} Specific heat $C$ vs $T$ in zero field and various applied fields along $[110]$. Inset: Temperature dependence of the peaks as a function of field. The line is a guide to the eye. Specific heat data from Ref. \citenum{Matsuhira09} on La$_2$Zr$_2$O$_7$ were subtracted to determine the value of the peak temperature. }
\end{figure}

Specific heat measurements show a broad peak around 2 K, in quantitative agreement with previous studies \cite{Matsuhira09, Kimura13} (see Figure \ref{Fig_CT}). This feature has been attributed to the development of a collective spin ice state. It should be noted however that the shape is quite different from canonical spin ices \cite{Ramirez99, Kimura13}. In addition, the peak temperature (about 2.2 K) is larger than the range of exchange interactions that can be inferred from magnetization measurements (which are a priori antiferromagnetic, contrary to the case of classical spin ice), which suggests that this anomaly may originate in another physical process, as will be discussed in section \ref{discuss}. 

When a magnetic field is applied along $[110]$, the amplitude of the peak increases, but its position is almost constant (actually, it seems to slightly move towards lower temperatures) for fields below 1 T.  At larger fields, the peak broadens and moves to larger temperatures. The field dependence of the peak is shown in the inset of Figure~\ref{Fig_CT}. For fields larger than 1 T, it can be reproduced by the linear equation $T_{\rm peak}({\rm K})=1.2+1.08\mu_0H({\rm T})$.

\subsection{Neutron diffraction} 
\label{diffraction}

To get more insight into the absence of quick saturation of the macroscopic magnetization, the field induced magnetic structures have been investigated by means of neutron diffraction up to 12~T. The data were collected using the D23 single crystal diffractometer (CEA-CRG, ILL France) operated with a copper monochromator and using $\lambda=1.28$ \AA. The field was applied successively along the $[1\bar{1}0]$ and $[111]$ direction. Refinements were carried out with the Fullprof software suite \cite{fullprof}.

\begin{figure}[t]
\includegraphics[width=8cm]{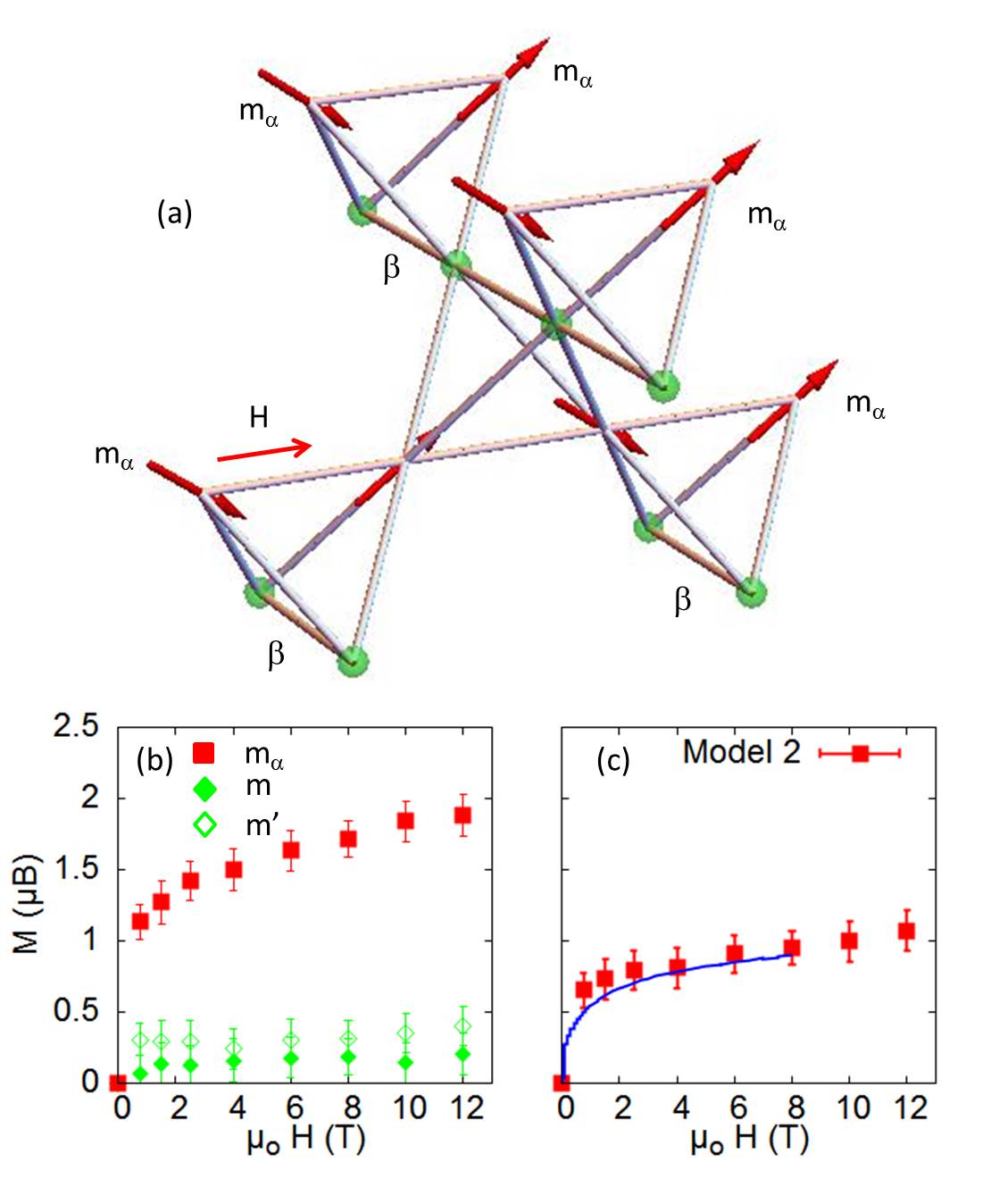}
\caption{\label{diffraction2}
(a) Sketch of the field induced structure for $H \parallel [1\bar{1}0]$ at T=50 mK. The red lines highlight the direction of the field. Green spheres illustrate the absence of magnetic moment. (b) Field dependence of the \pr\, ordered moments ($m_{\alpha}, m$ and $ m'$ defined in the text). (c) shows the magnetization calculated from diffraction results along with the macroscopic measurements at 90 mK (blue line). }
\end{figure}

When the field is applied along a $[1\bar{1}0]$ axis, the pyrochlore lattice splits into different sub-lattices, the so called $\alpha$ and $\beta$ chains, which are respectively parallel and perpendicular to the field direction, see Figure \ref{diffraction2}(a) and Table \ref{h1m10} (this nomenclature was introduced in Ref. \onlinecite{Hiroi03}). The local anisotropy axes $\vec{z}_i$ are respectively at 35 ($\vec{m}_{3,4}$) and 90 degrees ($\vec{m}_{1,2}$) of the applied field. 

In \hoti, \dyti\, and \tbti, neutron diffraction measurements \cite{Fennell05,Sazonov10,Sazonov11,Clancy09} have shown that the $\alpha$ moments align along their anisotropy axis with a net ferromagnetic component along the field. The $\beta$ chain moments adopt, however, different specific relative orientations described by a ${\bf k}= (0,0,1)$ propagation vector, giving rise to magnetic intensity on the ``forbidden" $Q$ vectors positions of the $Fd\bar{3}m$ space group. 

\begin{figure}[t]
\includegraphics[width=8cm]{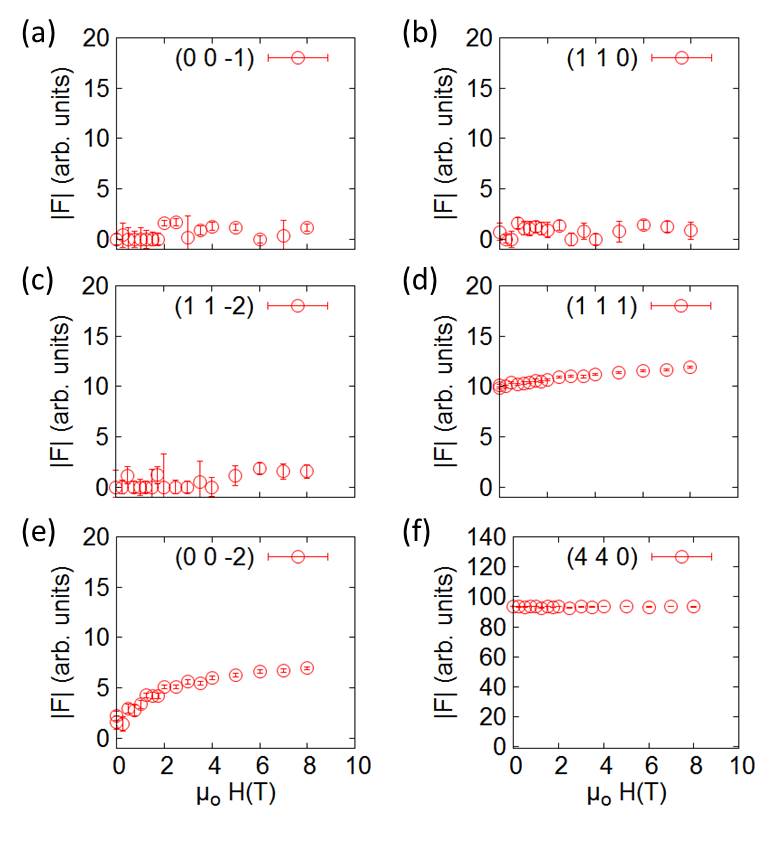}
\caption{\label{diffraction1} Field dependence of the structure factor obtained from neutron diffraction for selected Bragg peaks and measured at T=50 mK. The field is applied along $[1\bar{1}0]$. $(00\bar{1})$, $(110)$ and $(11\bar{2})$ are ``forbidden'' in the $Fd\bar{3}m$ space group and have essentially a zero intensity. The other ones are allowed and indeed have a significant intensity.}
\end{figure}

\begin{table}
\begin{tabularx}{\linewidth}{p{0.7cm}p{1.8cm}p{2.4cm}*{1}{>{\arraybackslash}X}}
\hline \hline
Site &  $\vec{z}_i$ & Model 1 & Model 2\\
1 ($\beta$) & $(1,1,-1)$    & $(0,0,0)$ & $(0,0,0) + m \vec{h}/h$\\
2 ($\beta$) & $(-1,-1,-1)$  & $(0,0,0)$ &  $(0,0,0) + m \vec{h}/h$ \\
3 ($\alpha$) & $(-1,1,1)$    & $ -m_{\alpha} \vec{z}_3 $ & $-m_{\alpha} \vec{z}_3 + m' \vec{h}/h$ \\
4 ($\alpha$) & $(1,-1,1)$   & $ +m_{\alpha} \vec{z}_4$ & $+m_{\alpha} \vec{z}_4 + m' \vec{h}/h$\\
\hline \\
Magnetization & &$m_{\alpha}/\sqrt{6}$ & $m_{\alpha}/\sqrt{6}+(m+m')/2$ \\
\hline \hline
\end{tabularx}
\caption{Direction of the magnetic moments in the different models discussed in the text for the magnetic field applied along $[1\bar{1}0]$.}
\label{h1m10}
\end{table}

In the present case of \przr, no additional peaks have been observed when ramping the field between 0 and 9~T. The intensity remains zero on the  ``forbidden" $Q$ vectors (see Figure \ref{diffraction1}a-c), which implies that the field induced structure is described by a ${\bf k}= (0,0,0)$ propagation vector. 
The refinement leads to the conclusion that the $\alpha$ moments behave as in conventional spin ices so that the corresponding ordered moment $m_{\alpha}=m_3=m_4$ increases with magnetic field (see Figure \ref{diffraction2}(a) and Table \ref{h1m10}, Model 1)
while, in contrast, along the $\beta$ chains (sites 1 and 2 in Table \ref{h1m10}), the ordered moment $m_{\beta}$ remains  essentially zero up to 12~T.
A slightly better fit is obtained by adding to this model additional components parallel to the applied field, $m \vec{h}/h$ and $m' \vec{h}/h$ for $\alpha$ and $\beta$ sites respectively (see Table \ref{h1m10}, Model 2). Both remain small, of the order of 0.2 $\mu_{\rm B}$. They involve the rise of transverse components with respect to the local anisotropy axis, which are induced by a mixing with the excited CEF levels due to the applied magnetic field. It is worth noting that their order of magnitude is consistent with recent calculations of the CEF \cite{Bonville} taking into account the complete basis of 4f states and not restricted to the ground spin-orbit multiplet of \pr\, ($^3H_4$). 
As shown in Figure \ref{diffraction2}(b), $m_{\alpha}$ struggles to grow and never saturates, even at 12~T. The calculated magnetization based upon this field induced structure smoothly increases with increasing field, in good agreement with the macroscopic magnetization reproduced as a blue curve in Figure \ref{diffraction2}(c). 

\begin{figure}[t]
\includegraphics[width=8cm]{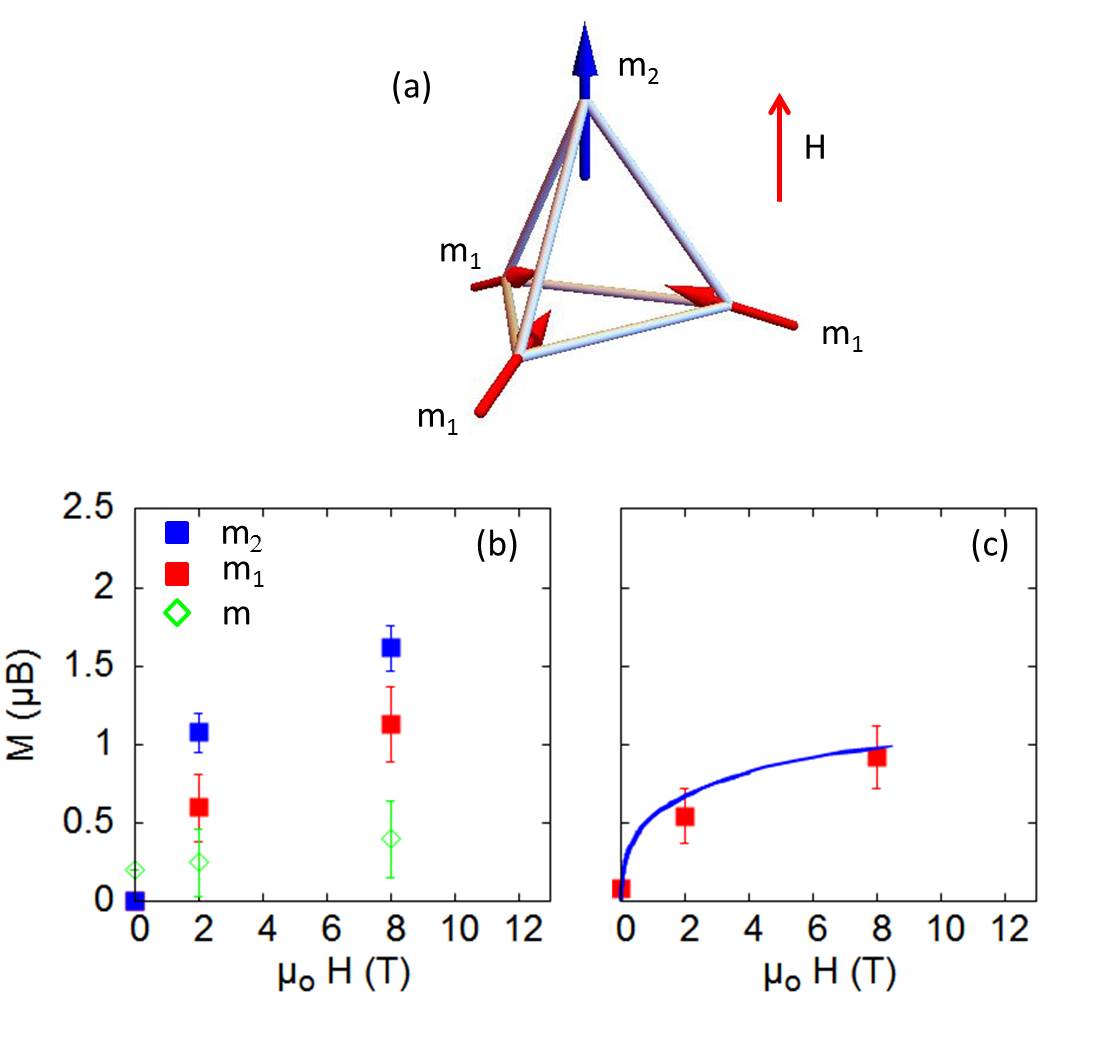}
\caption{\label{diffraction3}
(a) Sketch of the ``1-out$-$3-in'' structure. (b) Field dependence for $H \parallel [111]$ of the magnetic moments $m_{1,3,4}$, $m_2$ and $m$ based on diffraction and magnetization results obtained at T=50 and 90 mK respectively (see also Table \ref{h111}). (c) shows the calculated magnetization along with the macroscopic measurements (blue line). }
\end{figure}

When the field is applied along the $[111]$ axis, the field induced structure can also be described by a ${\bf k}= (0,0,0)$ propagation vector. In that case, one should distinguish $\vec{m}_2$, which has its anisotropy axis along the field, from the three left moments $\vec{m}_{1,3,4}$ that are at 71 degrees off (or 109 depending on their direction). From the diffraction data only, we could not refine a unique magnetic structure. We thus chose to constrain the magnetic moments to match the magnetization obtained in macroscopic measurements. This leads to a structure which resembles the ``1-out$-$3-in'' structure (see Figure \ref{diffraction3}(a)) except that $\vec{m}_2$ and $\vec{m}_{1,3,4}$ have different amplitudes. In addition, a component of 0.2 $\mu_B$ parallel to the field, similar to what has been obtained when $H \parallel [1\bar{1}0]$, is needed (see Figure \ref{diffraction3}(b) and Table \ref{h111}). The calculated magnetization based upon this field induced structure is shown in Figure \ref{diffraction3}(c). 
Importantly, for both magnetic field directions, the diffraction data confirm that the system hardly magnetizes as a function of field.

\begin{table}
\begin{tabularx}{\linewidth}{p{0.7cm}p{2.4cm}*{1}{>{\arraybackslash}X}}
\hline \hline
Site & $\vec{z}_i$ & Model \\
1 & $(1,1,-1)$    & $+m_1 \vec{z}_1+m \vec{h}/h$\\
2 & $(-1,-1,-1)$  & $ -m_2 \vec{z}_2$ \\
3 & $(-1,1,1)$    & $+m_1 \vec{z}_3+m \vec{h}/h$ \\
4 & $(1,-1,1)$    & $+m_1 \vec{z}_4+m \vec{h}/h$\\
\hline \\
Magnetization & & $(m_1+m_2)/4 + 3m/4$ \\
\hline \hline
\end{tabularx}
\caption{Direction of the magnetic moments for the magnetic field applied along $[111]$.}
\label{h111}
\end{table}


\subsection{Spin dynamics} 
\label{spin_dynamics}

We finally investigate the spin dynamics, both in zero and applied field, that emerge from these ground states (note that we study here the very low energy response, well below the first CEF level located at 10 meV). To this end, inelastic neutron scattering experiments were conducted at low temperature $T=60$ mK on a large \przr\, single crystal (Figure \ref{photo-pr}) mounted in order to have the $(hh0)$ and $(00\ell)$ reciprocal directions in the horizontal scattering plane. The sample was attached to the cold finger of a dilution insert, and the magnetic field was applied along $[1\bar{1}0]$. Time of flight measurements were carried out on the IN5 spectrometer operated by the Institut Laue Langevin (France). A wavelength $\lambda=4.9$ \AA\ was used yielding an energy resolution of about 80 $\mu$eV. The data have been processed with the {\it Horace} software \cite{horace}, transforming the time of flight, sample rotation and scattering angle into $\omega$ energy transfer and $Q$-vectors. We then took constant energy slices and constant $Q$ cuts in $(Q,\omega)$ space to show respectively the $Q$ and energy-dependence of the response. The integration range around a given $(Q, \omega)$ point was $(\Delta h, \Delta \ell , \Delta \omega)$ with $\Delta h=\Delta \ell=0.05$ and $\Delta \omega=$ 0.1 meV ($h$ and $\ell$ are in reduced reciprocal lattice units). The rather large value of $\Delta \omega$, roughly the energy resolution, was chosen to offer a better statistics. Triple axis measurements (TAS) were also carried out at the 4F2 cold spectrometer installed at LLB (France). We used a final wave-vector $k_f= 1.2$~\AA$^{-1}$, leading again to an energy resolution of about 80 $\mu$eV.

\begin{figure*}[t]
\includegraphics[height=8cm]{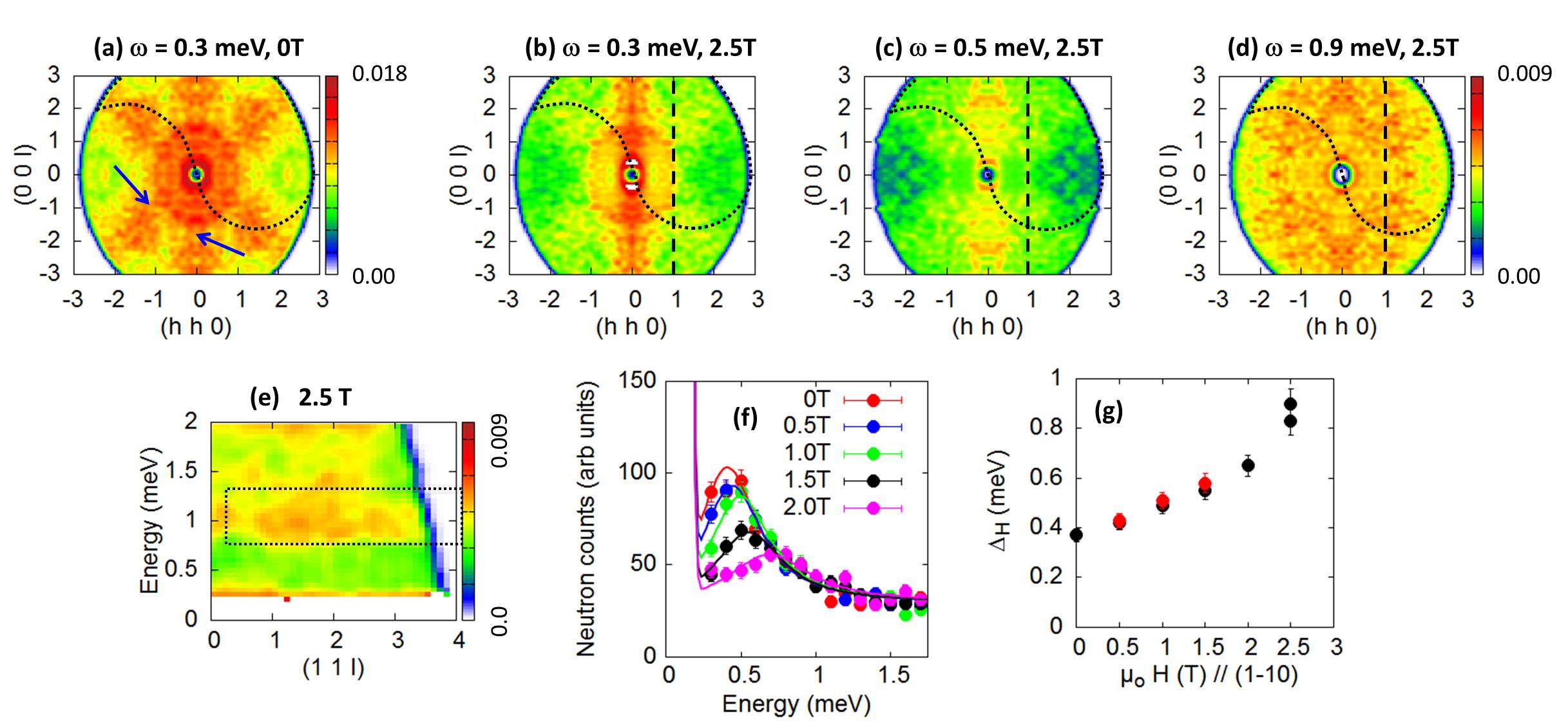}
\caption{\label{INS2} Inelastic neutron data at 60 mK. 
(a) $Q$-vector map in zero field of the inelastic neutron intensity at 0.3~meV. Blue arrows indicate the pinch point positions. The black dotted lines delineate the area actually measured, and the data have been symetrized.
(b-d) $Q$-vector maps of the inelastic neutron intensity at 0.3, 0.5, and 0.9 meV and 2.5 T applied along $[1\bar{1}0]$. The dashed line corresponds to the direction of $Q$ in map (e).
(e) $(\omega,Q=(11\ell))$ map measured at 2.5 T showing the flat dispersionless mode at about 0.9 meV and highlighted by the dashed rectangle. 
(f) Representative triple axis spectra carried out at $Q=(1,1,1)$. The lines are fits according to a Lorentzian profile, showing a strong mode at the energy $\Delta_H$.
(g) Field dependence of $\Delta_H$; the red and black points correspond to the experimental data at $Q=(1,1,1.5)$ and $(1, 1, 1)$ respectively. 
}
\end{figure*}

In zero applied magnetic field, the present data show that the spin dynamics consist in a broad low energy response whose structure factor resembles the specific pattern observed in classical spin-ice, with arm like features along the $(00\ell)$ and $(hhh)$ directions. This is illustrated in Figure \ref{INS2}(a) which shows a slice taken at $\omega=0.3$~meV. The $Q$-width of the signal is obviously smaller at the pinch point positions $(002)$ and $(111)$ (labeled with blue arrows). Turning now to the energy dependence of the response, the TAS data (see Figure \ref{INS2}(f)) can be accounted for by a Lorentzian profile describing an overdamped mode at the characteristic energy $\Delta$ with a lifetime $1/\Gamma$:
\begin{eqnarray}
\label{lorentzian}
I(Q,\omega) & = & \frac{A}{1-e^{-\omega/T}} \times  \nonumber \\
& & \left(\frac{\Gamma}{(\omega-\Delta)^2+\Gamma^2}-\frac{\Gamma}{(\omega+\Delta)^2+\Gamma^2}\right) 
\label{fitneutron}
\end{eqnarray}
We find $\Delta \approx \Gamma \approx$ 0.4 meV. This mode can be compared to the discrete excitation measured at low temperature in \prhf \cite{Sibille16} and centered at $\Delta \approx 0.2$~meV, as well as to the profile observed in \prsn \cite{Zhou08}. The broadening in the case of \przr\ could be due to chemical inhomogeneities or disorder \cite{Foronda15, Duijn05}.

It should be stressed that our results are consistent with the INS data reported by Kimura et al \cite{Kimura13}. Our experiments especially confirm that the spectrum is {\it mostly inelastic}. In Ref. \onlinecite{Kimura13}, the elastic scattering is estimated to be 10\% of the total response, and we note that according to their energy resolution (0.12 meV), it cannot be excluded that at least part of this very weak elastic response might come from the inelastic channel. In our experiments, any elastic contribution, if it exists, could not be detected, because of the large elastic incoherent background of the cryomagnet.

New information is obtained from INS results performed under a magnetic field applied along the $[1\bar{1}0]$ axis. The response encompasses a first contribution visible at low energies. A slice taken at 0.3 meV and 2.5 T, presented in Figure \ref{INS2}(b), displays a single arm along $(00\ell)$. Some intensity is visible along $(hhh)$ but strongly weakened compared to zero field (note that the color scales of (a) and (b-d) are different in Figure \ref{INS2}). This resembles much the rod like diffuse scattering observed in \hoti \cite{Clancy09} under an applied field, except that the signal is inelastic in the case of \przr. 
No spin wave dispersion could be detected from these data, perhaps because of the weakness of the signal. With increasing the energy transfer $\omega$, the slice shown in Figure \ref{INS2}(c) shows that the intensity of the arm feature along $(00\ell)$ progressively weakens.
As explained in section II, owing to the non-Kramers nature of the \pr\ ion, the {\it inelastic} rod like signal observed at 2.5 T suggests that the ground state of these moments is quadrupolar. The specific $Q$ dependence (rod-like) denotes that the magnetic excitations built above the quadrupolar state are formed within the $\beta$ chains.
This picture is consistent with the diffraction data obtained for $H \parallel [1\bar{1}0]$ (Section \ref{diffraction}) showing the lack of elastic response at the Bragg positions and that would have indicated a long range order of magnetic moments (as in \hoti).

Interestingly, with further increase of the energy transfer, a second contribution arises, which takes the form of a dispersionless mode at $\omega=\Delta_H$. This character is illustrated in Figure \ref{INS2}(e). It displays an intensity map taken as a function of energy and wave-vector along $(1,1,\ell)$ at 2.5 T. Here, the mode appears as a roughly flat and broad excitation at a characteristic energy $\Delta_H \approx 0.9$~meV. 
To the accuracy of the experiment, the intensity of the mode does not depend on $Q$ (see Figure \ref{INS2}(d)). TAS measurements show that this mode emerges from the zero field broad response for fields as small as 0.5 T. This is illlustrated in Figure \ref{INS2}(f) which fetaures spectra taken at $Q=(1,1,1)$ for various fields. Fitting the data through the Lorentzian profile (Equation \ref{lorentzian}), we find that the characteristic energy $\Delta_H$ strengthens upon increasing field, as shown in Figure \ref{INS2}(g). Concomitantly, the amplitude weakens while the damping increases. Interestingly, $\Delta_H$ shows a similar field dependence as the peak temperature of the specific heat (see Figure \ref{Fig_CT}), suggesting that the two phenomena are likely connected.


\section{Discussion}
\label{discuss}

\subsection{Role of quadrupolar degrees of freedom}

As described above (Section \ref{spin_dynamics} and in Ref. \onlinecite{Kimura13}), the zero field neutron scattering signal is essentially inelastic. It can be described by a flat mode, whose width might be induced by inhomogeneities in the sample. This observation reminds the case of the kagom\'e antiferromagnet KFe$_3$(OH)$_6$(SO$_4$)$_2$ \cite{Matan06}, and more recently the pyrochlore system \ndzr \cite{Petit16}, in which an inelastic flat mode was interpreted as a zero energy mode (the kagom\'e weather vane mode and the spin ice pattern respectively) lifted up to finite energy by an additional term in the Hamiltonian (a Dzyaloshinskii-Moriya term and an octopolar term respectively). 

In \przr, the quadrupolar degrees of freedom, which are expected to play an important role \cite{Onoda10}, could be the key ingredient to explain this flat mode at finite energy. Indeed, the \pr\, ion is a non-Kramers ion. As discussed in Section \ref{sectionnk}, the presence of an inelastic signal can thus be interpreted as the signature that the main components of the pseudo spins lie, in the ground state, within the {\it local} $xy$ plane, and not in the magnetic $z$ direction. This would correspond to a quadrupolar ground state, from which magnetic excitations emerge and are revealed through the inelastic signal. In that context, the dynamical rod-like signal observed at 2.5 T when $H\parallel [1 \bar{1} 0]$ can be interpreted as magnetic fluctuations emerging from the state formed by the quadrupolar moments within the $\beta$ chains. 

This proposal is consistent with the shape of the measured magnetization curves. When a field is applied the magnetization increases much more slowly than what would be expected for classical Ising spins in presence of small antiferromagnetic interactions. This smooth increase can be understood as a competition between the magnetic field and the quadrupolar correlations: the magnetic field component along the local $\vec{z}$ axis promotes the rise of magnetic moments to the detriment of the quadrupoles. 

In that picture, the broad peak observed in the specific heat would involve the quadrupolar degrees of freedom. It is worth noting that the description of the specific heat in terms of monopoles is hard to reconcile with the energy ranges present in the system: the temperature of the specific heat anomaly (about 2 K) is larger than the Curie-Weiss temperature ($|\theta_{\rm{CW}}|<1$~K) characterizing the magnetic interaction range. The specific heat anomaly temperature is especially larger than the ``canonical'' spin ice (\hoti\, and \dyti) one, despite a larger Curie-Weiss temperature in these systems\cite{Gardner10}. In addition the negative Curie-Weiss temperature in \przr\, suggests antiferromagnetic interactions, in contrast with the spin ice description which calls for positive ${\cal J}^{zz}$ interactions. 

\subsection{Input of the mean field approximation}
\label{meanfield}

To go a step further, and understand qualitatively how these quadrupoles might be correlated, we now examine the Hamiltonian Eq (\ref{h}) at the mean field level. The spin dynamics is calculated in the RPA, a method that has been developed at length in the context of pyrochlore magnets \cite{Jensen,kao03,petit14,robert15}. 

\begin{figure}[t]
\includegraphics[width=6cm]{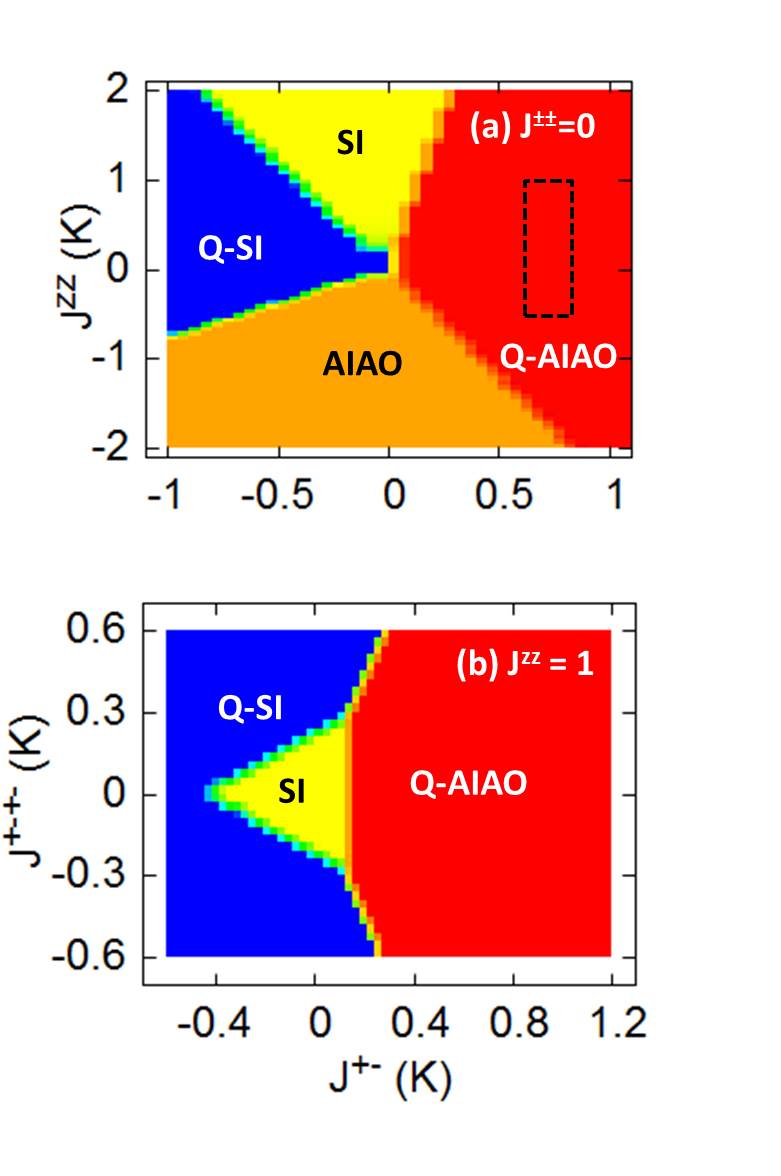}
\caption{\label{diagphase} Mean field phase diagram of the model defined by Eq. (\ref{h2}). The SI and AIAO phases correspond respectively to the ordered spin ice phase and to the ``all-in$-$all-out'' antiferromagnetic phase. Both of them are magnetic, with pseudo spins ordered along the local $\vec{z}$ axes. In the Q-AIAO phase, the pseudo spins are ordered and parallel to the same (symmetry equivalent) local axis within the $xy$ plane (``ferro-pseudo spin order''). In this sense, this phase is characterized by an antiferro-quadrupolar order. The Q-SI phase is characterized by a ferroquadrupolar order. In (a), ${\cal J}^{\pm\pm}=0$, while in (b), ${\cal J}^{zz}=1$. The dashed rectangle shows the region of interest for \przr.}
\end{figure}

\begin{figure}[t]
\includegraphics[width=9cm]{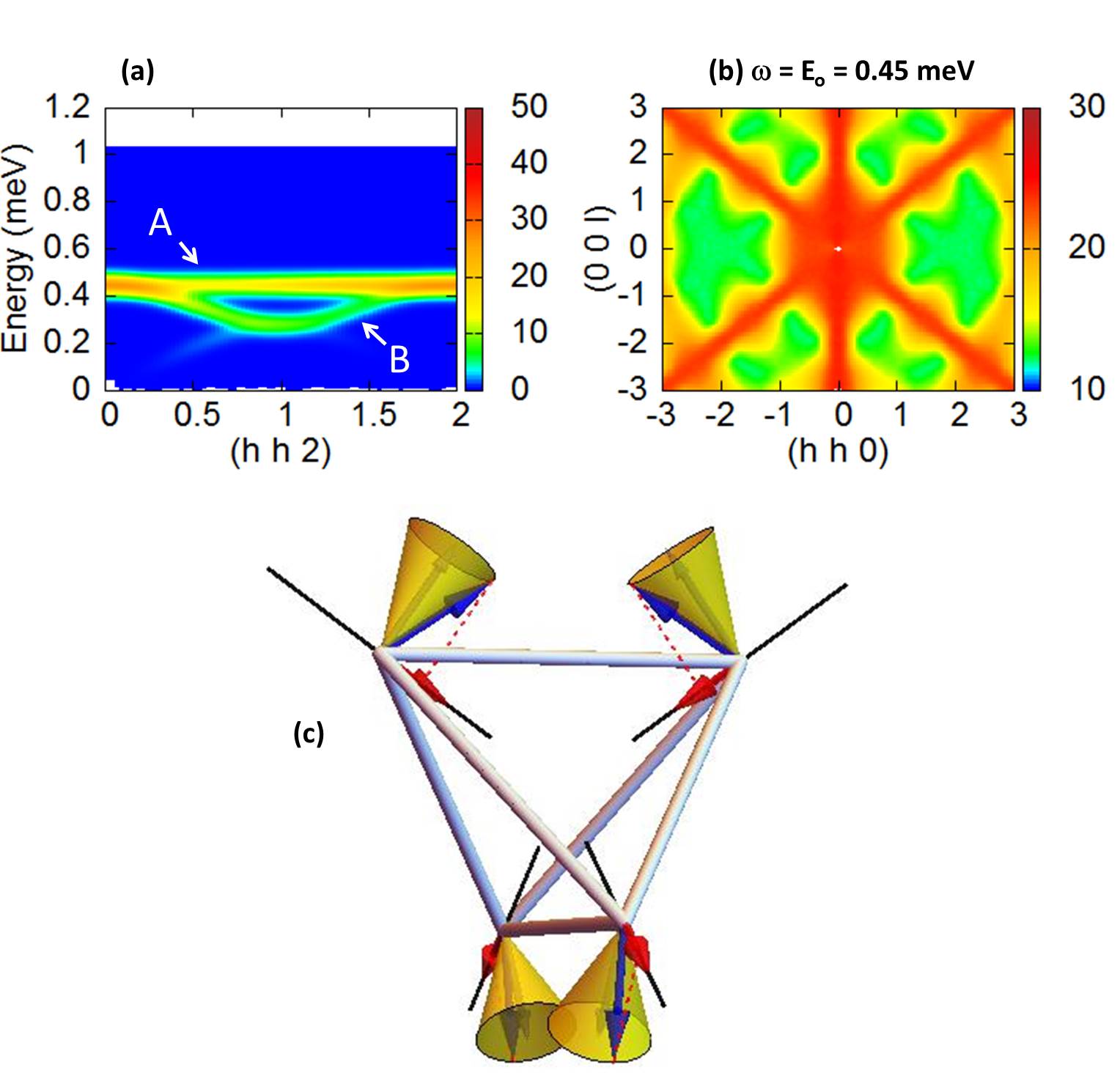}
\caption{\label{MF-RPA-fin1} Spin dynamics calculated in the RPA. The ground state is the antiferro-quadrupolar phase Q-AIAO, characterized by an ordering of the pseudo-spin $\sigma$ along the $x$ local axes. (a) $(\omega,Q=(hh2))$ map calculated for ${\cal J}^{\pm}$=0.7 K and ${\cal J}^{zz}$=-0.5 K showing the presence of the dispersionless mode at $E_o=0.45$ meV (labeled by an ``A"). (b) Zero field $Q$-vector map taken at $E_o$. (c) Precession of the pseudo spins (illustrated by cones) in the dispersionless mode. The blue arrows feature a snapshot of the relative orientation of the pseudo-spins. Projecting those pseudo spins along the $\vec{z}$ axis directions (red arrows) gives two projections pointing into and two out of the center of the tetrahedron. As a function of time, the spins oscillate in a manner that fullfils the ``2-in$-$2-out'' ice rule.
}
\end{figure}

\begin{figure*}[t]
\includegraphics[height=8cm]{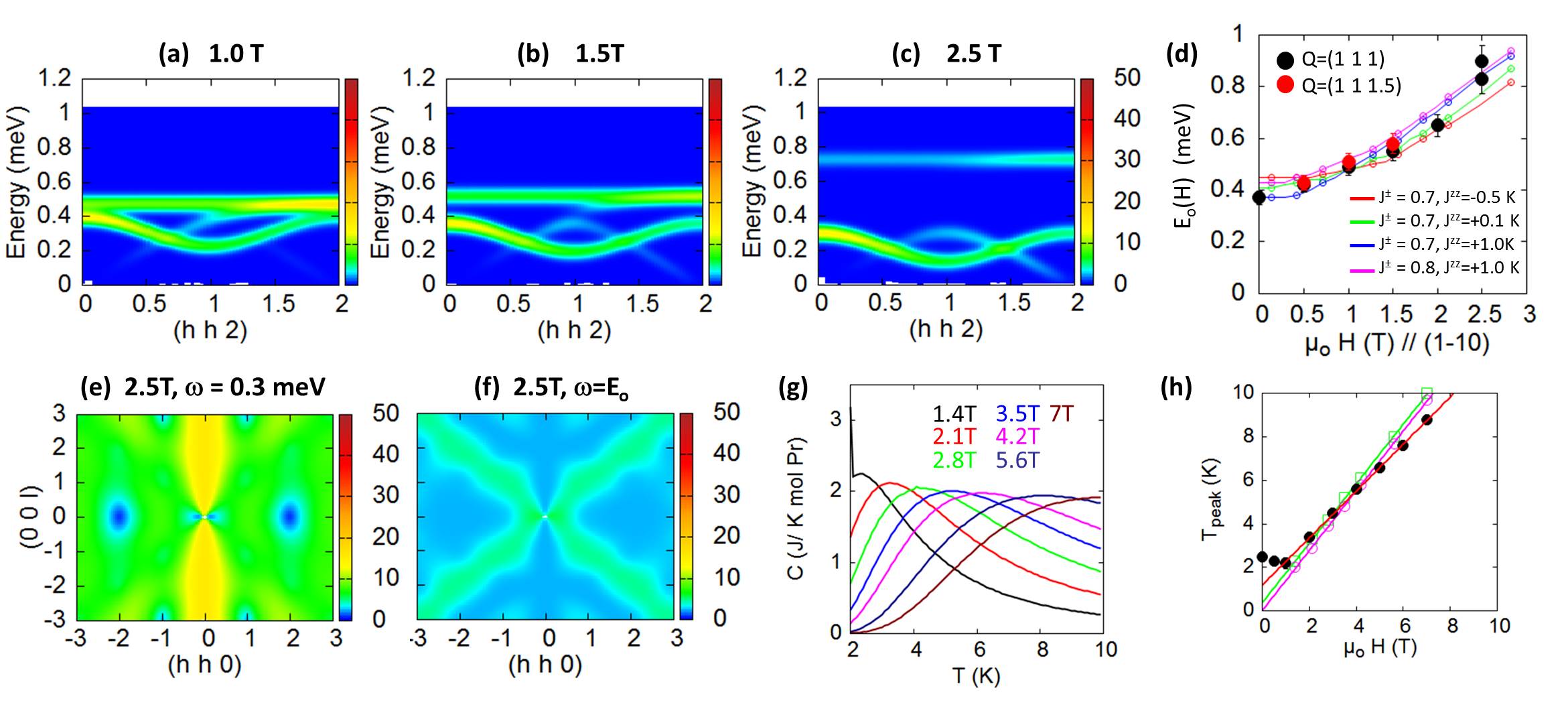}
\caption{\label{MF-RPA-fin2} Spin dynamics calculated in the RPA for ${\cal J}^{\pm}$=0.7 K and ${\cal J}^{zz}$=-0.5 K and under a field applied along $[1\bar{1}0]$. (a-c): $(\omega,Q=(hh2))$ maps at different fields. Note that the energy of the mode at $E_o$ progressively increases with increasing field. (d) Field dependence of $E_o$. The lines are results from RPA calculations with ${\cal J}^{\pm}$=0.7 K and ${\cal J}^{zz}$=-0.5 K (red line), ${\cal J}^{zz}$=0.1 K (green line) and ${\cal J}^{zz}$=1 K (blue line) and ${\cal J}^{\pm}$=0.8, ${\cal J}^{zz}$=1 K (magenta line). The red and black points correspond to the experimental $\Delta_H$ at $Q=(1,1,1.5)$ and $(1,1,1)$ respectively. 
(e) Map of the energy integrated response up to $\omega=0.3$ meV at 2.5 T. It nicely compares with Figure \ref{INS2}(b).
(f) Map taken at 2.5 T and $\omega=E_o$, to compare with the experimental data shown in Figure \ref{INS2}(c). 
(g) Calculated specific heat above the transition towards the ordered Q-AIAO state. It features a Schottky-like anomaly, similar to what is found in experiment (see Figure \ref{Fig_CT}). (h) shows the field dependence of the temperature at which the specific heat is maximum. The magenta line is a linear fit to the calculations performed with ${\cal J}^{\pm}$=0.8 K and ${\cal J}^{zz}$=1 K. The green one was calculated  with ${\cal J}^{\pm}$=0.7 K and ${\cal J}^{zz}$=0.1 K. The red line is the fit of the experimental data (black points) presented in the inset of Figure \ref{Fig_CT}.}
\end{figure*}

\subsubsection{Phase diagram}

We first look at the phase diagram computed as a function of ${\cal J}^{zz}$, ${\cal J}^{\pm}$ and ${\cal J}^{\pm\pm}$ (Figure \ref{diagphase}). In agreement with Ref. \onlinecite{Onoda11}, four different phases are obtained: an antiferromagnetic ``all-in$-$all-out'' phase (AIAO), a ferromagnetic ``2-in$-$2-out'' ordered spin ice phase (SI) and two quadrupolar phases (denoted with a ``Q" prefix). It is worth noting that the ordered SI phase obtained at this level of approximation is replaced by the classical spin ice for ${\cal{J}^{\pm\pm}}=0, {\cal{J}^{\pm}}=0$, and by a U(1) spin liquid phase in more elaborate theories \cite{Lee12}. Both quadrupolar phases correspond to an ordering of the pseudo spin $\sigma$ within the $xy$ plane ($\theta_i=\pi/2$). They carry a zero magnetic moment and have either the ``spin-ice" nature, with alternate directions of $\sigma$
, or an AIAO nature (the pseudo spins point along the same local direction
). In the latter case, the mean field approximation leads to an ordered phase, but owing to the $xy$ symmetry, it is likely that it remains disordered in more elaborate approaches. Note that the present Q-AIAO and Q-SI quadrupolar phases are the mean field variants of the ``antiferroquadrupolar'' and ``ferroquadrupolar'' Higgs phases of Ref. \onlinecite{Lee12} (yet the boundaries between the different phases are slightly different). 

\subsubsection{Spin dynamics in the Q-AIAO phase }

The Q-AIAO phase is particularly relevant for our purpose. Throughout this phase only (our calculations are restricted to ${\cal J}^{\pm\pm}$=0 for simplicity), the RPA spin dynamics consist in a dispersionless excitation at an energy $E_o$ (labeled with an { ``A" in Figure \ref{MF-RPA-fin1}(a)), whose neutron structure factor is the spin-ice pattern (see Figure \ref{MF-RPA-fin1}(b)). Analytical calculations based on a spin wave expansion around the Q-AIAO order allow one to better understand the physical essence of this dispersionless mode. We find that it corresponds to a precession of the pseudo spins at a frequency $E_o/\hbar$ around their equilibrium direction with:
\begin{equation}
E_o  = 4\sqrt{ {\cal J}^{\pm} ( 3 {\cal J}^{\pm} -{\cal J}^{zz}/2)}
\label{eo}
\end{equation}
The eigenvectors of this mode are such that in each tetrahedron, the four spins can be divided into two pairs, characterized by a phase shift of $\pi$ (see also the Appendix \ref{dispersionlessmode}). For instance, the dynamical magnetization on the summits of a tetrahedron can be written as:
\begin{eqnarray*}
\vec{m}_{1,2}(t) & = & g_{\parallel} \sigma \cos{(E_o/\hbar~t)}~\vec{z}_{1,2} \\
\vec{m}_{3,4}(t) & = & g_{\parallel} \sigma \cos{(E_o/\hbar~t+\pi)}~\vec{z}_{3,4}
\end{eqnarray*}
which is nothing but the ``2-in$-$2-out" ice rule. It also can be understood as a dynamical divergent free magnetization, hence leading to the spin-ice dynamical structure factor. Figure \ref{MF-RPA-fin1}(c) shows a sketch of the relative orientations of the pseudo spin in this particular mode. Finally, we observe that $E_o$ goes to zero at the boundary with the SI phase (see Appendix \ref{appendix_QAIAO}). 

The RPA also reveals collective excitations (labeled with a ``B" in Figure \ref{MF-RPA-fin1}(a)). Their dispersion lies below or above the flat mode depending on the values of the parameters (see Appendix \ref{appendix_QAIAO}). With decreasing ${\cal J}^{zz}$ (becoming stongly negative), these dispersing branches go soft at the Bragg positions of the AIAO phase, signaling the phase transition towards this magnetic state. 

The spectra and the spin-ice pattern shown in Figure \ref{MF-RPA-fin1} have been obtained for ${\cal J}^{\pm}=0.7$~K, ${\cal J}^{zz}=-0.5$~K. These parameters have been chosen so that $E_o$ corresponds to the experiment energy scale $\Delta$ (see below).

When a magnetic field is applied along $[1\bar{1}0]$, our calculations carried out in the Q-AIAO phase show that a static magnetic moment on the $\alpha$ sites is restored, while the $\beta$ sites remain quadrupolar in nature, in agreement with what we have observed in neutron diffraction. In addition, the  energy of the dispersionless mode increases with increasing the field (see Figure \ref{MF-RPA-fin2}(a-c)) and its structure factor  becomes less featured, as illustrated in Figure \ref{MF-RPA-fin2}(f). Concomitantly, the characteristic energy of the dispersing branches softens. Integrating this low energy part of the response up to 0.3~meV gives the map shown in Figure \ref{MF-RPA-fin2}(e) and characterized by a single arm along $(00\ell)$. It is worth noting the close correspondence with the experimental data reported in Figure \ref{INS2}(b). 

We also determined the temperature and magnetic field dependence of the magnetic specific heat. The latter was computed above the transition towards the ordered Q-AIAO state. It shows a maximum, similar to what is found in experiment. We find that this maximum shifts linearly to higher temperature with increasing field, as shown in Figure \ref{MF-RPA-fin2}(g). 

Finally, we have calculated the susceptibility and the magnetization ($M$ vs $H$) curves. In constrast with experiment, in presence of quadrupolar terms, the susceptibility saturates when decreasing the temperature, and remains smaller than the measured one. The quadrupolar terms slow down the increase of the magnetization with magnetic field compared to a model whithout these terms, making the calculated curves closer to the experimental ones. The latter are however smoother and the saturation values are smaller. This might be partly explained by the mixing with excited states of the crystal field in presence of magnetic field which tends to decrease the effective moment, and cannot be taken into account in such pseudo-spin $1/2$ approach (with or without quadrupolar terms).

\subsubsection{Proposal}

The above mean-field approach shows that the $E_o$ mode can be induced in presence of a positive ${\cal J}^{\pm}$ coupling between the $xy$ components of the pseudo spins. This occurs provided that ${\cal J}^{\pm}$ is strong enough with respect to the magnetic exchange ${\cal J}^{zz}$, precluding the stabilization of the conventional SI and AIAO magnetic phases (the mean field energy of the Q-AIAO is $-6{\cal J}^{\pm}$ to be compared with $-2{\cal J}^{zz}$ which is the energy of the SI phase).

Based on these results, we propose that the mode observed at $\Delta$ in \przr\, can be interpreted in terms of the dynamical spin-ice mode at $E_o$ of the Q-AIAO phase. The data in presence of a magnetic field are consistent with this proposal, suggesting that $\Delta_H$ follows the field dependence of $E_o$. 

To estimate a range of coupling parameters of the \przr\, Hamiltonian that would qualitatively describe the experimental observations, a systematic exploration of the Q-AIAO phase has been carried out, assuming however ${\cal J}^{\pm\pm}=0$ for the sake of simplicity. We determined numerically the field induced structure, the spin dynamics, especially the field dependence of $E_o$ (see Figure \ref{MF-RPA-fin2}(d)), and calculated the instantaneous magnetic correlations by integrating this spectrum over the energy. We also determined the temperature and magnetic field dependence of the magnetic specific heat (see Figure \ref{MF-RPA-fin2}(h)). This systematic survey of the Q-AIAO phase yields a good qualitative agreement with the experimental data for:
\begin{eqnarray*}
0.7 \leq {\cal J}^{\pm} \leq 0.8 \textrm{ K} \\
-0.5 \leq {\cal J}^{zz} \leq 1 \textrm{ K}
\end{eqnarray*}
along with ${\cal J}^{\pm\pm}=0$ which was our initial simplifying assumption. 

These parameters are quite different from the ones proposed in Ref. \onlinecite{Onoda10,Onoda11}, which tentatively locate \przr\, in the Q-SI phase. With a negative value of ${\cal J}^{\pm}$, however, the spin-spin correlation function does not display the ice-like pattern (see Figure 11 in this Ref. \onlinecite{Onoda11}), in contradiction with experiments. 

Our calculations with the above parameters confirm that, in presence of quadrupolar interactions, a spin ice pattern can be obtained despite a negative ${\cal J}^{zz}$, which is usually expected to stabilize an AIAO phase. This pattern is, however, shifted in the inelastic channel. This picture where quadrupolar degrees of freedom are at play thus resolves the apparent contradiction between the negative Curie-Weiss temperature, suggesting antiferromagnetic interactions, and the spin ice like structure factor observed in neutron scattering. 
 
Nevertheless, no transition towards a quadrupolar ordered state, predicted in this mean-field approach, is observed in specific heat which suggests that the ground state of \przr\, is rather a quadrupolar liquid with correlations typical of the Q-AIAO phase. 
In addition, the low temperature susceptibility behavior suggests that additional fluctuations between the quadrupolar and magnetic components have to exist in the ground state, so that the moment is not purely quadrupolar even at very low temperature, and which may prevent the quadrupolar ordering.
The spin-ice mode at $E_o$ appears strongly broadened in the experiments, maybe due to these fluctuations but likely also because of inhomogeneities. From the structure of the mean field equations (see Eq. (\ref{h2})), we anticipate that a strain field such that $v_i \equiv v \leq 0$ for all sites would spread the values of $E_o$, accounting for a significant broadening.

\section{Conclusion}

We have performed a detailed study of the properties of the quantum spin ice candidate \przr\, using macroscopic and neutron scattering measurements. In particular, magnetization and diffraction measurements show that the system hardly magnetizes at very low temperature. ${\bf k}=0$ field induced structures are obtained when the field is applied along the $[1\bar{1}0]$ and $[111]$ directions. Along $[1\bar{1}0]$, the magnetization and diffraction data are consistent with a structure where the ordered moment is carried by the so called $\alpha$ chains only. Along $[111]$, we find a ``1-out$-$3-in" structure with moments of different amplitude. For both directions, the spins align along their local anisotropy axis with however a small transverse component.

The specific heat measurements show that above 1 T, the broad anomaly reported in Ref. \onlinecite{Matsuhira09, Kimura13} shifts to larger temperatures. Our inelastic scattering measurements show that the spectrum can be viewed as a broad flat mode centered at about 0.4 meV with a magnetic structure factor which resembles the spin ice pattern. These data confirm that the response is mostly dynamical \cite{Kimura13}. When a magnetic field is applied along $[1\bar{1}0]$ (at least up to 2.5 T), the $Q$-structure of the response at low energy changes to a rod-like pattern, similar to what was observed in \hoti \cite{Clancy09}. In addition, a well defined mode forms, whose energy increases when the field increases, in the same way as the temperature of the specific heat anomaly, and which is featureless in $Q$ at 2.5 T. 

This set of experiments can be qualitatively understood by introducing a coupling between quadrupolar degrees of freedom in the Hamiltonian widely accepted for pyrochlores magnets. These terms lift the ``spin ice" diffuse pattern up to finite energy. Using a mean-field approach that takes into account these quadrupolar terms, we show that the field induced behavior can be qualitatively understood, and propose a set of exchange parameters able to account qualitatively for the data in this approximation. Our analysis points out that the ground state of \przr\ might support antiferroquadrupolar correlations \cite{Lee12,Onoda10,Onoda11}, from which emerge magnetic ice-like excitations. 

Phenomenologically, we propose that \przr\ could be described as a quadrupole liquid, characterized by short-range Q-AIAO correlations. The spin ice like excitations are shifted to finite energy, highlighting the fact that the quadrupolar state is ``protected'' from the spin-ice state. The fact that pinch points may exist in the elastic channel \cite{Kimura13} as well as the low temperature behavior of the magnetic susceptibility suggest that some magnetic moments can re-form to the detriment of the quadrupolar state. In this picture, the actual ground state would consist of an assembly of both quadrupoles and magnetic moments, i.e. to a state characterized by fluctuations between the quadrupolar liquid with Q-AIAO correlations and the spin ice phase. The dispersionless mode would probably broaden in energy, acquiring a finite lifetime, so that the pinch points would also exist at zero energy.  Further theoretical studies, beyond the mean-field approach, are thus needed to give a more complete picture of the \przr\ ground state and analyze quantitatively our observations.


\acknowledgments
MCH and GB acknowledge financial support from the EPSRC, United Kingdom, Grant No. EP/M028771/1.

\appendix

\section{Crystal electric field}
\label{appendix_CF}

The CEF coefficient determined in Ref. \onlinecite{Bonville} are reproduced in Table \ref{bnm}
\begin{table}[h]
\begin{tabularx}{\linewidth}{p{0.5cm}*{10}{c}>{\centering\arraybackslash}X}
\hline
\hline
$J_z$ & $-4$ & $-3$ & $-2$ & $-1$ & $0$ & $1$ & $2$ & $3$ & $4$ \\ \hline
& $a$ & & & $b$ & & & $c$ &  &  \\ \hline
$| \uparrow\rangle$    & 0.894 & 0 & 0 & $0.448$ & 0 & 0 & -0.024 & 0 & 0  \\
$| \downarrow\rangle$ & 0 & 0 & -0.024 & 0 & 0 & -0.448 & 0 & 0 & 0.894 \\
$| 1 \rangle$ & 0 & 0.299 & 0 & 0 & -0.909 & 0 & 0 & -0.299 & 0 \\
\hline \hline
\end{tabularx}
\caption{Ground state wave functions of \pr\ in \przr. The Wybourne coefficients (in $\mu$eV) reproduced from Ref. \citenum{Bonville} are $B_{20}=-631$, $B_{40}=-32.36$, $B_{43}=-467.4$, $B_{60}=0.245$, $ B_{63}=1.464$ and $B_{66}=-1.907$. }
\label{bnm}
\end{table}
With these values, one obtains the Land\'e factors $g_{\parallel} = 5.5$ and $g_{\perp} =0$. CEF levels are found at 10, 57, 82, 93 and 109 meV.

\section{Spin-spin correlation function}
\label{appendix_Sq}

Let us write formally the dynamical spin-spin correlation function $S(Q,\omega)$ measured by neutron scattering, in terms of the actual eigenstates $|\Phi_n \rangle$ with energies $E_n$ (above the ground state): 
\begin{eqnarray*}
S(Q,\omega) =  \sum_{i,j} e^{iQ(R_i-R_j)} \sum_{n,m} \frac{e^{-E_{n}/k_B T}}{Z} \\
\times \langle \Phi_{n} | \vec{J}_{\perp,i} | \Phi_{m} \rangle \langle \Phi_{m} | \vec{J}_{\perp,j} | \Phi_{n} \rangle \delta(\omega-E_n+E_m)
\end{eqnarray*}
with $Z =\sum_{n} \exp\left({-E_{n}/k_B T}\right)$ and where the symbol $\perp$ indicates that one must consider the components perpendicular to the scattering wavevector $Q$. At low temperature, keeping the ground and first excited state, this reduces to:
\begin{eqnarray*}
S(Q,\omega) \approx  |\langle \Phi_{G} | \sum_i e^{iQR_i} \vec{J}_{\perp,i} | \Phi_{G} \rangle |^2 \delta(\omega) \\
+ |\langle \Phi_{1} | \sum_i e^{iQR_i} \vec{J}_{\perp,i} | \Phi_{G} \rangle |^2 \delta(\omega-E_1)
\end{eqnarray*}
hence to an elastic contribution at $\omega=0$, and an inelastic one at $\omega=E_1$. 

In a classical picture, the ground state $|\Phi_G \rangle$ of the above Hamiltonian (\ref{h2}) can be described as a state where on each site of the pyrochlore lattice, the expectation value of the pseudo spin $\vec{\sigma}=(\sigma_x,\sigma_y, \sigma_z)$ is oriented in the direction specified by {\it local} spherical angles $\theta_i$ and $\phi_i$ (see Figure \ref{Fig1}): $\theta_i$ defines the polar angle relative to the local CEF axes 
and $\phi_i$ is the angle within the $xy$ plane:
\begin{equation*}
\left\{
\begin{array}{ccc}
|\Phi_G \rangle & = & | \phi_{G,1}~...~\phi_{G,i}~ ...~ \phi_{G,N} \rangle \\
|\phi_{G,i} \rangle & = &  \cos \frac{\theta_i}{2} | \uparrow \rangle_i + e^{i\phi_i} \sin \frac{\theta_i}{2} | \downarrow \rangle_i
\end{array}
\right.
\end{equation*}
where $N$ is the (infinite) number of sites. Those angles depend on the Hamiltonian. As expected for instance in the RPA or spin wave approximation, the lowest energy excited states should contain one flip of the pseudo spin, possibly delocalized over the lattice. $|\Phi_1 \rangle$ is thus constructed as:
\begin{equation*}
|\Phi_1 \rangle = \sum_i  ~ C_i~| \phi_{G,1}~...~\phi_{1,i}~ ...~ \phi_{G,N} \rangle
\end{equation*}
where $|\phi_{1,i}\rangle$ describes such a flip of the pseudo spin $\sigma$ at site $i$. The values of the $C_i$ coefficients depend on the Hamiltonian and remain to be determined. Written in the $|\uparrow,\downarrow \rangle$ subspace, $|\phi_{1,i}\rangle$ must be normalized and orthogonal to $|\phi_{G,i}\rangle$, and thus of the form:
\begin{equation*}
|\phi_{1,i} \rangle =  -e^{-i\phi_i} \sin \frac{\theta_i}{2} | \uparrow \rangle_i + \cos \frac{\theta_i}{2} | \downarrow \rangle_i
\end{equation*}
The relevant matrix elements then write (in the global coordinates):
\begin{eqnarray*}
\langle \phi_{G,i} | \vec{J}_i |  \phi_{G,i}\rangle & = & \mu \cos \theta_i~\vec{z}_i \\
\langle \phi_{1,i} | \vec{J}_i | \phi_{G,i} \rangle  & = & -\mu ~ e^{i\phi_i} ~\sin \theta_i~\vec{z}_i
\end{eqnarray*}
leading to the following elastic and inelastic contributions:
\begin{eqnarray*}
S(Q,\omega=0)    &\approx & \mu^2  |\sum_{i} e^{iQ R_i} \cos \theta_i ~\vec{z}_{\perp,i} |^2 \\
S(Q,\omega=E_1) &\approx & \mu^2   |\sum_{i} C_i~e^{iQ R_i} ~ e^{i\phi_i} \sin \theta_i ~\vec{z}_{\perp,i} |^2
\end{eqnarray*}

\section{Analysis of the neutron diffraction data}
\label{appendix_diff}

As explained in the main text, we ramped the field on various $Q$ position between 0 and 9~T (see Figure \ref{diffraction1-total}). We observed that the neutron intensity remains zero on the ``forbidden'' peaks of the $Fd\bar{3}m$ space group. This implies that the field induced structure is described by a ${\bf k}= (0,0,0)$ propagation vector.

The analysis of the neutron diffraction data has then two stages. First, high temperature (10K) data have been recorded and fitted using the Fd$\bar{3}$m space group. The free parameters of the fit were the scale factor, the position of the oxygen, the isothermal and the extinction coefficients. The low temperature data have then been fitted via a model containing both the crystalline and ${\bf k}= (0,0,0)$ magnetic structures. Yet the parameters of the crystalline structure were fixed to the values obtained at 10K. For the data obtained with $H \parallel [111]$, the fit was carried out considering the magnetic structure only and using the difference between the neutron intensities at 10K and at low temperature. 

\begin{figure}[t]
\includegraphics[height=12cm]{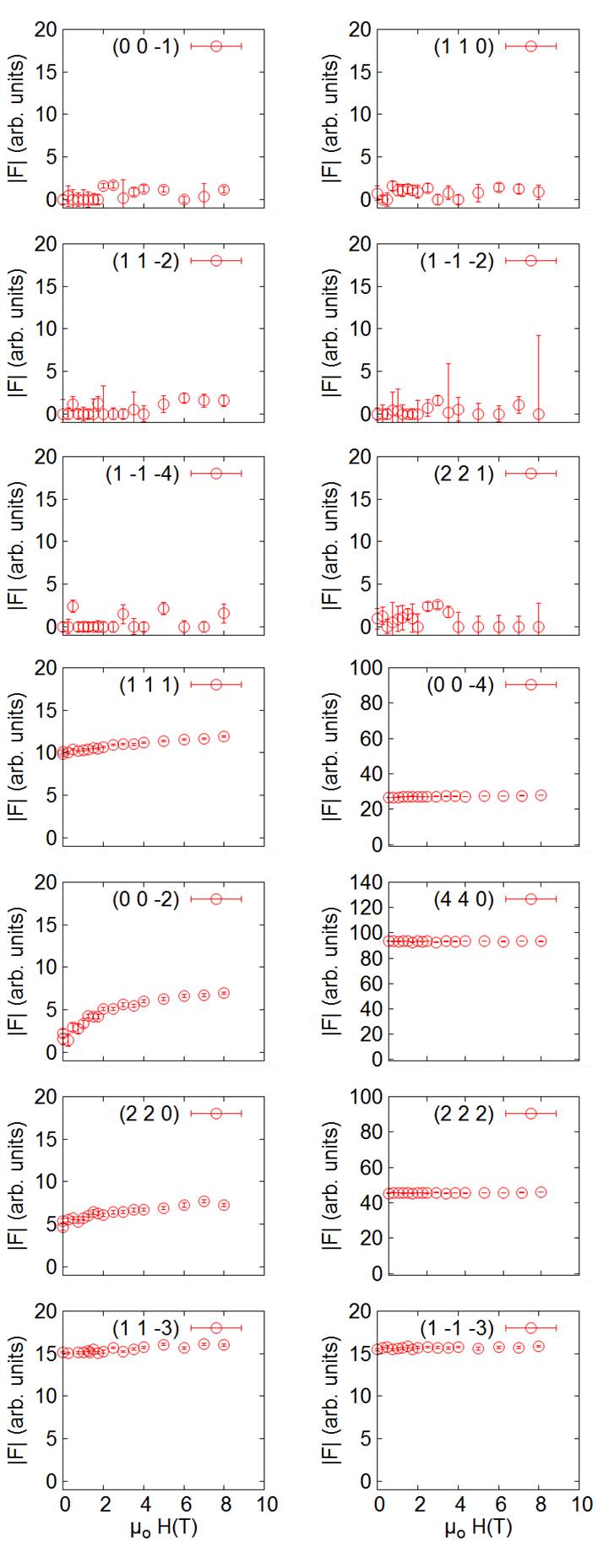}
\caption{\label{diffraction1-total} Field dependence of the structure factor obtained from neutron diffraction for various Bragg peaks. The field is applied along $[1\bar{1}0]$. The six upper Bragg positions are forbidden in the Fd$\bar{3}$m space group and have essentially a zero intensity. The eight lower ones are allowed and indeed have a significant intensity.}
\end{figure}

\section{Evolution of the spin dynamics in Q-AIAO phase}
\label{appendix_QAIAO}
In this section we illustrate in Figure \ref{MF-RPA-2} the evolution of the spin dynamics calculated within the RPA in the Q-AIAO phase. As explained above, the spin excitation spectrum encompasses a flat mode at $E_o$ together with dispersive branches below or above $E_o$. We observe that $E_o$ goes soft as the border with the SI phase is approached i.e. with increasing ${\cal J}^{zz}$ or decreasing ${\cal J}^{\pm}$. In contrast, with decreasing ${\cal J}^{zz}$, the dispersing branches go soft at the Bragg positions of the AIAO phase, signaling the phase transition towards this magnetic state.  

\begin{figure*}[t]
\includegraphics[height=8cm]{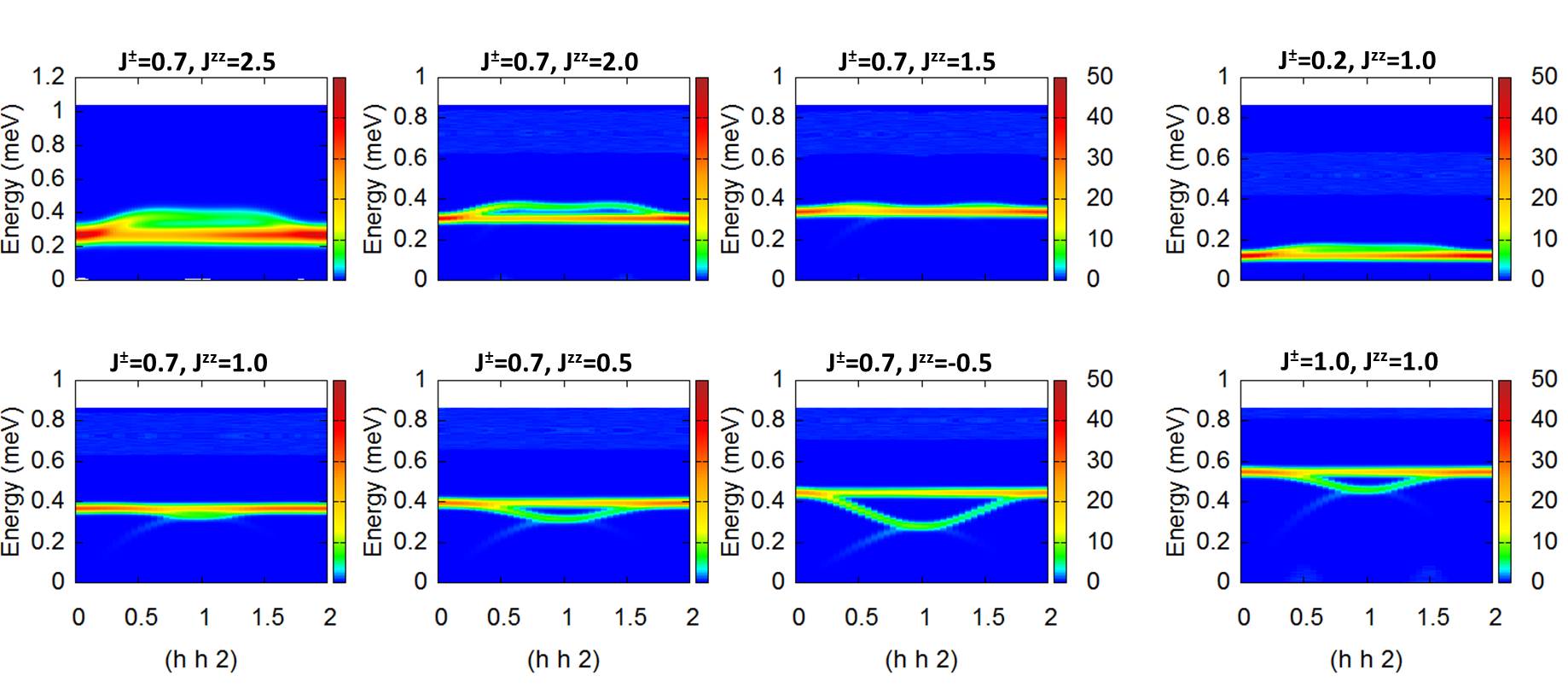}
\caption{\label{MF-RPA-2} Spin dynamics calculated within the RPA in the Q-AIAO phase. The spectra are shown along $(hh2)$ for various sets of parameters.}
\end{figure*}

\section{Dispersionless mode}
\label{dispersionlessmode}

\begin{figure}[t]
\includegraphics[width=5cm]{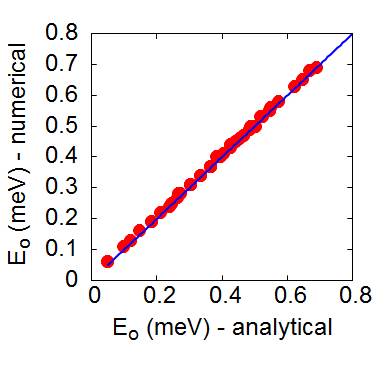}
\caption{\label{Eo-Cspe} Energy $E_o$ of the dispersionless mode within the Q-AIAO phase as a function of the parameters of the model; the analytical expression is given by Eq. (\ref{eo}).}
\end{figure}

To better understand the physical origin of the dispersionless mode, we proceed with analytical calculations on the basis of a spin wave expansion out of the Q-AIAO order. To this end, we introduce on each site $a_i^+$ and $a_i$ bosons that create or annihilate local deviations of the pseudo spin. The spin wave Hamiltonian writes \cite{Petit11}:
\begin{equation*}
{\cal H}  = {\bf a}^+~{\cal K}~{\bf a}
\end{equation*}
with ${\bf a}^+ = 
\left( a^+_1, a^+_2, ...a^+_i...  a^+_N, a_1, a_2, ...a_i... a_N \right)$
and ${\cal K}$ is a $2N \times 2N$ matrix :
\begin{eqnarray*}
{\cal K} & = &
\left(
 \begin{array}{cc}
-\sigma \Omega_i \delta_{i,j} + \frac{\sigma}{2} s_i J_{i,j} \bar{s}_j & +  \frac{\sigma}{2}s_i J_{i,j} s_j \\
 \frac{\sigma}{2} \bar{s}_i J_{i,j} \bar{s}_j & -\sigma\Omega_i \delta_{i,j} +  \frac{\sigma}{2}\bar{s}_i J_{i,j} s_j 
\end{array}
\right) \\
\Omega_i &=& \sum_{\ell}  R_{i,3} J_{i,\ell} \ R_{\ell,3}\\
s_i & = & R_{i,1} + iR_{i,2}
\end{eqnarray*}
where $R_i$ is a 3-column matrix $R_i=(R_{i,1}R_{i,2},R_{i,3})$ (see Table \ref{axes}), $J_{i,j}$ is the exchange matrix that couples the spins at sites $i$ and $j$. Using the Hamiltonian given by Eq. (\ref{h2}), the definition of the local axes, and owing to the pyrochlore structure, we find: 
\begin{eqnarray*}
\Omega_i              &=  & \Omega = -12 {\cal J}^{\pm} \\
s_i J_{i,j} \bar{s}_j & = & \epsilon_{i,j} \left(2{\cal J}^{\pm} -{\cal J}^{zz} \right) = \epsilon_{i,j}  A\\
s_i J_{i,j} s_j         & = &  \epsilon_{i,j} \left(2{\cal J}^{\pm} +{\cal J}^{zz} \right) = -\epsilon_{i,j} B
\end{eqnarray*}
with 
\begin{equation}
\epsilon_{i,j} =\pm 1
\label{eps}
\end{equation}
for neighboring $(i,j)$ spins (zero otherwise), and
\begin{equation}
\sum_{j\ne i, j \in \Delta_{i}} \epsilon_{i,j} = 1
\label{tetra}
\end{equation}
for each spin $i$ in a tetrahedron $\Delta_{i}$. With the convention of Table \ref{axes}, we have $\epsilon_{1,2}=\epsilon_{3,4}=-1, \epsilon_{1,3}=\epsilon_{1,4}=1, \epsilon_{2,3}=\epsilon_{2,4}=1$.

\begin{table}[h]
\begin{tabularx}{\linewidth}{p{2cm}*{2}{c}>{\centering\arraybackslash}X}
\hline
\hline
Site & $R_i$ \\
1
& 
$
\left(
\begin{array}{ccc}
-1/\sqrt{3} & -1/\sqrt{2} & -1/\sqrt{6} \\
-1/\sqrt{3} &  1/\sqrt{2} & -1/\sqrt{6} \\
1/\sqrt{3} & 0 & -2/\sqrt{6} \\
\end{array}
\right)
$
\\
2
&
$
\left(
\begin{array}{ccc}
1/\sqrt{3} &  1/\sqrt{2} &  1/\sqrt{6} \\
1/\sqrt{3} & -1/\sqrt{2} &  1/\sqrt{6} \\
1/\sqrt{3} & 0 & -2/\sqrt{6} \\
\end{array}
\right)
$ 
\\
3
&
$
\left(
\begin{array}{ccc}
-1/\sqrt{3} & -1/\sqrt{2} &  1/\sqrt{6} \\
1/\sqrt{3} & -1/\sqrt{2} & -1/\sqrt{6} \\
1/\sqrt{3} & 0 & 2/\sqrt{6} \\
\end{array}
\right)
$
\\
4 &
$
\left(
\begin{array}{ccc}
1/\sqrt{3} &  1/\sqrt{2} & -1/\sqrt{6} \\
-1/\sqrt{3} &  1/\sqrt{2} &  1/\sqrt{6} \\
1/\sqrt{3} & 0 & 2/\sqrt{6} \\
\end{array}
\right)
$ 
\\
\hline \hline
\end{tabularx}
\caption{Local axes in the pyrochlore lattice}
\label{axes}
\end{table}
The spin wave Hamiltonian is diagonalized by a Bogolubov transform which involves new bosons operators $\alpha$ and $\alpha^+$. The ground state of the model is then the vacuum of these operators. The energies of the spin waves and the associated eigenvectors $(..., u_i,...,..., v_i,...)$ must then be solution of: 
\begin{eqnarray*}
-\sigma \Omega u_i + \frac{\sigma}{2} \sum_j \left( s_i J_{i,j} \bar{s}_j u_j +   s_i J_{i,j} s_j  v_j \right) &=& E_o u_i \\
-\frac{\sigma}{2} \sum_j \left( \bar{s}_i J_{i,j} \bar{s}_j u_j + \bar{s}_i J_{i,j} s_j v_j \right) + \sigma\Omega v_i &=& E_o v_i 
\end {eqnarray*}
hence:
\begin{eqnarray*}
-\sigma \Omega u_i + \frac{\sigma}{2} \sum_j \left( A\epsilon_{i,j} u_j  -B\epsilon_{i,j} v_j \right) &=& E_o u_i \\
-\frac{\sigma}{2} \sum_j \left( -B \epsilon_{i,j} u_j + A \epsilon_{i,j} v_j \right) + \sigma\Omega v_i &=& E_o v_i 
\end {eqnarray*}
Taking advantage of \ref{tetra}, we now look for a particular solution where in each tetrahedron $\Delta_i$ : 
\begin{equation}
\sum_{j \in \Delta_i} \epsilon_{i,j} u_j = u_i,~~\sum_{i \in \Delta_i}  \epsilon_{i,j} v_j = v_i
\label{divf}
\end{equation}
Since each site belongs to two tetrahedra, we obtain: 
\begin{eqnarray*}
-\sigma \Omega u_i + 2 \frac{\sigma}{2} A u_i -  2 \frac{\sigma}{2}B v_i &=& E_o u_i \\
2 \frac{\sigma}{2}B u_i + \sigma\Omega_i v_i - 2 \frac{\sigma}{2}Av_i &=& E_o v_i 
\end {eqnarray*}
Solving for $E_o$, we find a solution which is independent of $i$ and thus corresponds to a dispersionless mode:
\begin{equation}
E_o = 2\sigma \times 4 \sqrt{{\cal J}^{\pm}(3 {\cal J}^{\pm}-{\cal J}^{zz}/2)}
\end{equation}
Note that an exhaustive survey of the Q-AIAO phase by numerical calculations confirms this analytic formula, as shown in Figure \ref{Eo-Cspe}. 

Eq. (\ref{divf}) defines the structure of the associated eigenvectors. Since the $u$ and $v$'s are identical on each site, the spins rotate in phase {\it within their local basis} at a frequency $E_o$ around the equilibrium direction. We proceed by calculating the spin at site $i$; it is the projection of the pseudo-spin along the CEF axes (redefined above as $R_{i,1}$):  
\begin{eqnarray*}
\vec{S}_i & = &  (g_{\parallel} \vec{R}_{i,1}.\vec{\sigma}_i) \vec{R}_{i,1} \\
\vec{\sigma}_i & = & \frac{g_{\parallel} \sqrt{2\sigma}}{2} \left(\bar{s}_i a_i + s_i a^+_i\right) + g_{\parallel} R_{i,3} (\sigma-a^+_i a_i) 
\end{eqnarray*}
Hence:
\begin{equation*}
\vec{S}_i = \frac{g_{\parallel}\sqrt{2\sigma}}{2} \left( a_i +  a^+_i\right) \vec{R}_{i,1} 
\end{equation*}
The contribution of the dispersionless modes to the spin-spin correlation function (at $\omega=E_o$) then writes:
\begin{eqnarray*}
{\cal S}(Q,E_o) & = & g_{\parallel}^2\sigma \sum_{i,j} e^{iQ (R_{i}-R_j)} (u_i +v_i)(u_j + v_j) \vec{R}_{i,1}.\vec{R}_{j,1} \\
 & = & g_{\parallel}^2\sigma (u+v)^2 |\sum_{i} e^{iQ R_{i}}  \vec{R}_{i,1}|^2 
\end{eqnarray*}
Owing, to the definition of the $R_{i,1}$ given in Table \ref{axes}, ${\cal S}(Q,E_o)$ has the same structure as the spin-ice pattern defined in section \ref{sectionnk}. 


\end{document}